\documentclass[10pt]{article}

\usepackage[top=1in, bottom=1in, left=1.25in, right=1.25in]{geometry} % one-inch margins
\usepackage{amsthm}
\usepackage{thmtools,thm-restate}
\usepackage{amsmath}
\usepackage{amssymb}
\usepackage[]{mdframed}
\usepackage{graphicx}
\usepackage[round]{natbib}
\usepackage[plain,noend]{algorithm2e}
\usepackage[dvipsnames]{xcolor}
\usepackage{url} % not crucial - just used below for the URL 
\usepackage{comment}
\usepackage{arydshln}
\usepackage{enumitem}
\usepackage{bm}
\usepackage{multirow} % for vertically alignment
\usepackage{hyperref}

\newcommand{\abst}{\vert t \vert}
\newcommand{\LK}{\mathrm{LK}}
\newcommand{\lk}{\mathrm{lk}}
\newcommand{\igamma}{\mathrm{Inv.Gamma}}
\newcommand{\indep}{\perp\!\!\!\perp}

\newcommand{\diagOHf}{\mathrm{diag}^{\frac{1}{2}}}
\newcommand{\diagInvOH}{\mathrm{diag}^{-1/2}}
\newcommand{\diagInvOHf}{\mathrm{diag}^{-\frac{1}{2}}}

\newcommand{\sun}{\mathrm{SUN}}
\newcommand{\sut}{\mathrm{SUT}}
\newcommand{\sue}{\mathrm{SUE}}
\newcommand{\psun}{\mathrm{pSUN}}
\newcommand{\OmegaBar}{\bar{\Omega}}

\newcommand{\Qkol}{Q_{Kol}}

\newcommand{\qkol}{q_{Kol}}

\newcommand{\Logis}{\mathrm{Logis}}

\newcommand{\Vfix}{V_{fix}}

\newtheorem{definition}{Definition}

\begin{document}

%%%%%%%%%%%%%%%%%%%%%%%%%%%%%%%%%%%%%%%%%%%%%%%%%%%%%%%%%%%%%%%%%%%%%%%%%%%%%%%
\title{\bf An Extension of the Unified Skew-Normal Family of Distributions and its Application to Bayesian Binary Regression}
\author{Paolo Onorati \\
 Department of Statistical Sciences, University of Padua, Italy\\
 and \\
 Brunero Liseo \\
 MEMOTEF, Sapienza University of Rome, Italy}
\date{}
\maketitle

\bigskip
\begin{abstract}
We consider the Bayesian binary regression model and we introduce a new class of distributions, the Perturbed Unified Skew-Normal ($\psun$, henceforth), which generalizes the Unified Skew-Normal ($\sun$) class. We show that the new class is conjugate to any binary regression model, provided that the link function may be expressed as a scale mixture of Gaussian CDFs. We discuss in detail the popular logit case, and we show that, when a logistic regression model is combined with a Gaussian prior, posterior summaries such as cumulants and normalizing constants can easily be obtained through the use of an importance sampling approach, opening the way to straightforward variable selection procedures. For more general prior distributions, the proposed methodology is based on a simple Gibbs sampler algorithm. We also claim that, in the $p>n$ case, our proposal presents better performances - both in terms of mixing and accuracy - compared to the existing methods.

We illustrate the performance through several simulation studies and two data analyses. Supplementary Materials for this article, including the R package $\psun$, are available online.
 \end{abstract}

\noindent
{\it Keywords:} Importance Sampling; Kolmogorov Distribution; Logistic Regression; Scale Mixture of Gaussian Densities.
\vfill

\newpage

%%-----------------------------------------------------------------------------------------------
\section{Introduction}
\label{sec:intro}

Binary regression is among the most popular and routinely used statistical methods in applied sciences. Standard Bayesian approaches and off-the-shelf packages are available today, see for example the R suites \texttt{brms} or \texttt{rstanarms}. However, there is no consensus on the choice of prior distributions and on how to discriminate among different link functions.
From a computational perspective, several approaches have been proposed and none of them seems to perform uniformly better than the others.

A remarkable contribution in this field has been provided by \citet{durante2019} who shows that the Unified Skew-Normal ($\sun$) family of densities \citep{arellano2006} can be used as a conjugate prior for the \textit{probit} regression model, that is the posterior distribution of the regression coefficients in a probit setting still belongs to the $\sun$ family. \citet{durante2019} himself provides methods for efficiently sampling from the $\sun$ distribution. His methodology is beneficial in the case of small sample sizes, if $n \le 100$, or at least when $p >> n$. The $\sun$ family of densities has been introduced by \citet{arellano2006}, who review and classify the various generalizations of the multivariate skew-normal distribution that have appeared in the literature.

Following this line of research, we introduce a larger class of distributions, namely the \textit{Perturbed SUN} ($\psun$ hereafter) family, which is obtained by replacing the two Gaussian laws appearing in the $\sun$ definition with two scale mixtures of Gaussian distributions. The double scaling produces a larger class of densities, which can be particularly useful in Bayesian binary regression. 

Even though our results are quite general, here we concentrate on logistic regression, which is by far the most popular regression model for binary outcomes in applied statistics. The mathematical background of our work is based on old results obtained by \citet{andrews1974} and \citet{stefanski1991}, who both showed that the logistic distribution can be represented as a scale mixture of Gaussian distributions when the mixing density is a particular transformation of a Kolmogorov distribution \citep{kolmogorov1933, smirnov1939}.

There have already been many attempts to apply a missing-data strategy - similar to the probit model - in the logit case; e.g. \citet{holmes2006, fruhwirth2010, gramacy2012}. \citet{polson2013}
exploit the representation of the logistic density in terms of a mixture of Gaussian densities with a Polya-Gamma mixing distribution. Their approach produces a useful method for generating values from the posterior distribution of the parameters in a logistic regression setup, via a Gibbs sampler. More recently, \citet{ultimate} have proposed a similar approach that can deal with a large class of models with a discrete response variable.

Our main contribution may be summarized as follows. Assume that the adopted link function can be expressed as a scale mixture of Gaussian distributions. Then:
\begin{enumerate}
\item We propose a general Gibbs algorithm that produces a posterior sample from the distribution of the coefficients of any binary regression model. We also show that the proposed approach does not suffer from the problem of poor mixing in the presence of unbalanced data.
\item Our algorithm works particularly well when $p >n$; in this case, at least using the logistic link, we show that our proposal outperforms the most popular alternative, the Polya-Gamma algorithm, in terms of {\it effective sample size per second}. 
\item If the prior distribution on the coefficients is Gaussian, we propose an {\it Importance Sampling} strategy which is extremely fast and allows us to easily produce both point and set estimates and an accurate estimate of the normalizing constant that opens the way to model and variable selection. We show its performance in the popular case of the logistic link.

\end{enumerate}

The rest of the paper is organized as follows. Section \ref{sec:pSUN} introduces the $\psun$ family and describes its properties. Section \ref{sec:LSBR} illustrates the use of the $\psun$ density as a conjugate prior for a large class of binary regression models. Section \ref{sec:logit} discusses the logistic regression setup in depth and introduces two alternative computational approaches, the former based on a Gibbs structure and the latter belonging to the importance sampling family. Section \ref{sec:exa} includes several comparative simulation studies and two applications with real data.
Section \ref{sec:conc} concludes.

%%--------------------------------------------------------------------------------------------------
\section{The Perturbed SUN Distribution} 
\label{sec:pSUN}

There have been, mainly at the beginning of this century, several alternative proposals for introducing some form of skewness to the multivariate normal distribution, starting with the seminal paper \citep{azz1996}. Most of the proposals were motivated by the need for more sophisticated models to cope with increasingly complex data structures; see for example \citet{genton-book}. Perturbations of the Gaussian density were also proposed in the form of more elaborated conjugate priors in \citet{ohagan-leo} and \citet{lilo}.
To compare and critically discuss the various extensions, \citet{arellano2006} introduced the Unified Skew-Normal ($\sun$) class of densities that includes many of the several proposals that have previously appeared in the literature. 
The SUN class of densities is based on the introduction of a fixed number $m$ of latent variables. A $p$-dimensional random vector $\beta$ is said to have a $\sun$ distribution, i.e.
$\beta \sim \sun_{p,m} \left( \tau, \Delta, \Gamma, \xi, \Omega \right)$
if its density function is 
\begin{equation*}
f_{\beta} (\beta) = \phi_\Omega(\beta-\xi) \frac{\Phi_{\Gamma - \Delta^\prime \bar{\Omega}^{-1} \Delta} \big( \tau + \Delta^\prime \bar{\Omega}^{-1} \mathrm{diag}^{-\frac12}(\Omega) (\beta-\xi) \big) }{\Phi_\Gamma(\tau)} \, , 
\end{equation*}
where $\xi \in \mathbb{R}^p$ is a location vector, $\tau \in \mathbb{R}^m,$ $\Gamma$ is an $m$-correlation matrix, $\Omega$ is a $p$-covariance matrix, and $\bar{\Omega} = \textrm{diag}^{-\frac12}(\Omega)\, \Omega\, \textrm{diag}^{-\frac12}(\Omega)$ is the corresponding correlation matrix. Finally, $\Delta$ is a $p \times m$ matrix, and $\phi_\Omega(\cdot)$ and $\Phi_\Omega(\cdot)$ denote the density and the cumulative distribution function of a centered Gaussian random variable with variance-covariance matrix $\Omega$, respectively.
In simple words, leaving aside the location and scale parameters, the SUN is the conditional distribution of a zero mean multivariate normal vector $Z$, given that another (correlated) normal vector $U$ is such that $U > -\tau.$
The above expression provides a stochastic representation of a $\sun$ random vector, as $\beta = \xi + \textrm{diag}^{1/2}(\Omega) Z \vert U + \tau > 0 $, where 
the constraint $U + \tau >0$ must be intended as component-wise, and 
\begin{equation*}
	\begin{bmatrix}
		Z \\ 
		U \\
	\end{bmatrix}
	\sim N_{p+m} \left( 
	\begin{bmatrix}
		0 \\ 
		0 \\
	\end{bmatrix}, 
	\begin{bmatrix}
		\bar{\Omega} & \Delta \\ 
		\Delta^\prime & \Gamma \\
	\end{bmatrix}
	\right) \, ;
\end{equation*}
notice that the general form of a $\sun$ distribution is not identifiable under permutations of the latent variable vector $U$ \citep{wang2023}.

We introduce a generalization of the above family by assuming that $Z$ is a scale mixture of Gaussian distributions and $U$ is a linear combination of $Z$ and another random vector distributed as a scale mixture of Gaussian densities. 
We introduce here a slightly different parametrization of the $\sun$ family, namely 
\begin{equation}
	\label{eq:sunSR}
	\beta = \xi + \mathrm{diag}^{\frac 12}(\Omega) Z \vert  T \leq A Z + b,
\end{equation}
where $A \in \mathbb{R}^{p \times m}, b \in \mathbb{R}^{m}$. 
In this new parametrization, the conditioning latent vector $T$
is independent of $Z$, and $ T \sim N_m(0, \Theta)$, with 
\begin{align*}
	\Theta =& \, \mathrm{diag}^{-\frac 12} \left( \Gamma-\Delta^\prime \bar{\Omega}^{-1} \Delta \right) \left( \Gamma-\Delta^\prime \bar{\Omega}^{-1} \Delta \right) \mathrm{diag}^{-\frac 12} \left( \Gamma-\Delta^\prime \bar{\Omega}^{-1} \Delta \right) \, , \\
	A =& \, \mathrm{diag}^{-\frac 12} \left( \Gamma-\Delta^\prime \bar{\Omega}^{-1} \Delta \right)\Delta^\prime\bar{\Omega}^{-1}\, , \\
 	b =& \, \mathrm{diag}^{-\frac 12} \left( \Gamma-\Delta^\prime \bar{\Omega}^{-1} \Delta \right) \tau \, .
\end{align*} 

\noindent
Here the parameter $\Delta$ must be such that  
$\Gamma-\Delta^\prime \bar{\Omega}^{-1} \Delta$ is positive definite. 
When using this parametrization, we denote $\beta \sim \sun^{\ast}_{p,m}(\Theta, A, b, \xi, \Omega)$.
With a little abuse of notation, for a generic $p$-dimensional vector $H$, we denote with $\mathrm{diag}(H)$ the $p$-dimensional diagonal matrix with diagonal entries given by the components of $H$.
We now introduce the $\psun$ family:
    \begin{definition}
    A $p$-dimensional random vector $\beta$ belongs to the $\psun$ class of densities 
    \begin{equation*}
    \psun_{p,m}\left( Q_V, \Theta, A, b, Q_W, \Omega, \xi \right)
    \end{equation*}
    if it has the following stochastic representation  
    \begin{equation*}
    \beta = \xi +  \diagOHf (\Omega) Z \vert T \leq A Z + b 
    \end{equation*}
    where $Z$ and $T$ are independent and, conditionally on $(W,V)$, 
    \begin{equation}
    \label{eq:psunSR}
    Z  \sim N_p(0, \OmegaBar_W ) \indep 
    T \sim N_m(0, \Theta_V)\,;\quad \, 
    W \sim Q_W(\cdot) \indep  V \sim Q_V(\cdot).
    \end{equation}
    In the previous expression, $\xi$ is a location parameter, $\Omega$ is a scale matrix, and $\OmegaBar$ is the corresponding correlation matrix.
    In addition, $\Theta_V = \diagOHf(V)\Theta \, \diagOHf(V)$, $\OmegaBar_W = \diagOHf(W)\OmegaBar \, \diagOHf(W)$ while $Q_V(\cdot)$ and $Q_W(\cdot)$ are generic CDFs defined on the positive orthant.
\end{definition}    

As noted by a referee, the new proposed $\psun$ class is also a member of the Selection distribution family, as defined in \citet{arellano2006b}. 
A closed-form expression of a $\psun$ density is available; see expression \eqref{eq:psunDDF} below. 
The actual computation of the denominator in the density \eqref{eq:psunDDF} depends on the specific choices of $Q_W$ and $Q_V$. 
In section \ref{subsec:IS}, we describe a method to evaluate it, in some special but important cases, including the case when the $\psun$ density is the posterior obtained using a logistic link and a Gaussian prior.
Moreover, from a statistical perspective, the $\psun$ family is not identifiable, without imposing restrictions on the nature of $Q_V(\cdot)$ and $Q_W(\cdot)$.
For the sake of clarity, we will use the following notation:
\begin{align*}
\phi_{\Omega, Q_W} (x) &= \int_{\mathbb{R}^p} \prod_{i=1}^p \left({w_i^{-\frac 12}} \right) \phi_\Omega \left(
\mathrm{diag}^{-\frac 12} (w)\, x \right) dQ_W(w) \, , \\
\Phi_{\Omega, Q_W} (x) &= \int_{\mathbb{R}^p} \Phi_\Omega \left( \mathrm{diag}^{-\frac 12} (w)\, x \right) dQ_W(w) \, , \\
\Psi_{Q_V, \Theta, A, Q_W, \bar{\Omega}}(b) &= \mathrm{P} (T - A Z \le b) \, , \ T \sim \Phi_{\Theta, Q_V}(\cdot) \indep Z \sim \Phi_{\bar\Omega, Q_W}(\cdot) \, , \\
\widetilde{\Psi}_{Q_V, \Theta, A, Q_W, \bar{\Omega}} \left(b, k \right) &= \mathrm{P}(T - A \widetilde{Z}_k \le b) \, ,
\end{align*}
where $\widetilde{Z}_k$ is the $k$-tilted distribution of $Z$ \citep{siegmund1976}. It is then easy to obtain the density function of a generic $\psun$ random vector, say $f_\beta(\cdot)$; in addition, an expression for the moment-generating function (MGF) is available, provided that there exists the MGF of $Z$, say $M_Z(\cdot)$, as shown by the following theorem.
\begin{restatable}{theorem}{psun_DDFMGF}
	If $\beta\sim \psun_{p,m}\left( Q_V, \Theta, A, b, Q_W, \Omega, \xi \right)$ then the density function and the MGF of $\beta$ can be written as
	\begin{align}
		\mathrm{\emph{A}.} &\hspace{1.9mm} \label{eq:psunDDF}
		f_\beta(\beta) = 
		\phi_{\Omega, Q_W} (\beta - \xi) \frac{
			\Phi_{\Theta, Q_V} \left( A\, \mathrm{diag}^{-\frac 12}(\Omega)
			(\beta - \xi) + b \right)
		}{
			\Psi_{Q_V, \Theta, A, Q_W, \bar{\Omega}}(b)} \, , \\
		%%%%%%%%%%%%%%%%%%%%%%%%%%%%%%%%%%%%%%%%%%%%%%%%%%%%%%%%%%%%%%%%%%%%%%%%%%%
		\mathrm{\emph{B}.} &\hspace{1.9mm} \label{eq:psunMGF}
		M_\beta(u) = e^{u^\prime \xi}M_Z\left(\mathrm{diag}^\frac{1}{2}(\Omega)u\right) \frac{ \widetilde{\Psi}_{Q_V, \Theta, A, Q_W, \bar{\Omega}} \left(b,\mathrm{diag}^\frac{1}{2}(\Omega)u \right)}{\Psi_{Q_V, \Theta, A, Q_W, \bar{\Omega}}(b)} \, .
	\end{align}
\end{restatable}

\noindent
\textbf{Proof:}
See Supplementary Materials.

The denominator of expression \eqref{eq:psunDDF} is expensive to compute even for moderately large values of $m$. In Section \ref{subsec:IS}, we describe a practical solution to this problem, at least in the special situation when $Q_W(\cdot)$ is a point-mass distribution and $Q_V(\cdot)$ is the CDF of independent random variables distributed as $4$ times the square of a Kolmogorov distribution. In this case, the computation of the numerator of \eqref{eq:psunMGF} is also simpler, since $\widetilde{Z}_k$ becomes Gaussian with all means equal to $k$. We will see below that this corresponds to the case of a logistic regression model with a Gaussian prior.

We conclude this section by proposing an algorithm for simulating values from a 
$\psun$ distribution; the method is quite general and it can be suitably refined depending on the specific situation.

\subsection{Sampling from a \texorpdfstring{$\boldsymbol{\psun}$}{\psun} distribution} 
\label{subsec:psunRNG}
From the definition of $\psun$, it is apparent that if $\beta \sim \psun_{p,m}\left( Q_V, \Theta, A, b, Q_W, \Omega, \xi \right)$ then 
\begin{equation*}
\beta \vert W,V \sim \sun^*_{p,m}\left( \Theta, \diagInvOHf(V) A \, \diagOHf(W), \diagInvOHf(V) \, b, \, \xi, \diagOHf(W) \, \Omega \, \diagOHf(W) \right)\, .
\end{equation*}
Thus a possible strategy is to sample $\beta$ conditionally on $W$ and $V$; however, to do that, one must be able to sample from the conditional distribution of 
$(W, V)\vert T \leq AZ + b$, that is 
\begin{equation*}
	f_{W,V}(W,V \vert T \le AZ +b) = \frac{\Phi_{\Theta_V+A\bar\Omega_WA^\prime}(b)}{\Psi_{Q_V, \Theta, A, Q_W, \bar{\Omega}}(b)} f_W(W) f_V(V) \, .
\end{equation*}
This task is typically difficult to perform; here we adopt a more general technique based on the Gibbs sampler. Let $ \mathrm{TN}_m(\ell, u, \mu, \Sigma)$ denote an $m$-variate Gaussian random vector truncated at $\ell$ and $u$, with a mean vector $\mu$ and a variance-covariance matrix $\Sigma$. Suppose that, at iteration $t$, one has $\beta^{(t)}, Z^{(t)}, T^{(t)}, W^{(t)}, V^{(t)}$; the updating step is as follows:

\begin{mdframed}
    \begin{tabbing}
	   \qquad \enspace Sample $V^{(t+1)} \sim V \vert T = T^{(t)}$ \\
	   \qquad \enspace Sample $W^{(t+1)} \sim W \vert Z = Z^{(t)}$ \\
	   \qquad \enspace Sample 	$Z^{(t+1)}, T^{(t+1)} \sim Z,T \vert T \le AZ+b, W = W^{(t+1)}, V = V^{(t+1)}$ \\
	   \qquad \qquad Set $\varepsilon = T^{(t+1)} - AZ^{(t+1)}$ and $\Sigma_\varepsilon = \Theta_{V^{(t+1)}}+A\bar\Omega_{W^{(t+1)}} A^\prime$ \\
      \qquad 	\qquad   Sample $\varepsilon$ $\sim TN_m(-\infty, b, 0, \Sigma_\varepsilon)$ \\
	   \qquad \qquad Set $H_\mu = -\bar{\Omega}_{W^{(t+1)}}\, A^\prime \Sigma^{-1}_\varepsilon$ \\
	   \qquad \qquad Set $H_\Sigma = (I + H_\mu A^\prime) \,\bar{\Omega}_{W^{(t+1)}}$ \\
	   \qquad \qquad Sample $Z^{(t+1)} \sim N_d(H_\mu \varepsilon, H_\Sigma)$ \\
	   \qquad \qquad Set $T^{(t+1)} = AZ^{(t+1)} + \varepsilon$ \\
	   \qquad $\Longrightarrow \beta^{(t+1)} = \xi + \mathrm{diag}^{1/2}(\Omega)Z^{(t+1)}$
    \end{tabbing}
\end{mdframed}

There are two key aspects in the proposed scheme. First, one must be able to sample from a $\mathrm{TN}_m(-\infty, b, 0, \Sigma_\varepsilon)$; it is possible to simulate independent values from it by using the minimax exponentially tilted algorithm of \citet{botev2017}: specifically, to speed up computation, we have adapted some functions of the R package \texttt{TruncatedNormal} to our specific case. Nonetheless, Botev's algorithm performs quite well when the normalizing constant of the truncated Gaussian distribution is not too small and the performance degrades as the dimension increases; in general, it works well as long as $m \le 100$. For larger values of $m$, Botev's approach is reliable when the normalizing constant does not rapidly decay as $m$ gets larger, for example when the $\Sigma_\varepsilon$ components are weakly correlated.
The second key aspect is that one must be able to sample from the full conditional distributions $W \vert Z$ and $V \vert T$. This is not always easy, since it depends on the specific values of $\Theta, \bar{\Omega},$ and the form of $Q_W(\cdot)$ and $Q_V(\cdot)$. We will describe below those instances of Bayesian binary regression where this task is simple, including probit and logit regression with the most common priors.

%%--------------------------------------------------------------------------------------------------
\section{Bayesian Linear Symmetric Binary Regression}
\label{sec:LSBR}
Here we illustrate the use of the $\psun$ family of distributions as a conjugate prior for a large class of binary regression models. Consider a general version of the model as 
\begin{equation}
	\label{eq:BR}
	\begin{split}
		Y_i \vert p_i \overset {\textrm{ind}} \sim& Be(p_i) \, , \ i = 1,2, \dots, n \, , \\
		p_i =& \Lambda \big( \eta(X_i) \big) \, ,
	\end{split}
\end{equation}
where $\Lambda: \mathbb{R} \rightarrow [0,1]$ is a known link function, $\eta(\cdot)$ is a calibration function, 
and $X_i\in \mathbb{R}^p$ is the $i$-th row of the design matrix $X$. 
Typically $\Lambda(\cdot)$ is a univariate CDF whose corresponding density is symmetric about $0$, and $\eta(x)$ takes a simple linear form, $x^\prime \beta$; we refer to this case as the Linear Symmetric Binary Regression model (LSBR). Let $\Lambda_n(u) = \prod_{i=1}^n \Lambda(u_i)$, $u = (u_1, u_2, \dots, u_n) \in \mathbb{R}^n$ and, for $y \in \{0,1\}^n$, let $B_y = [2\, \mathrm{diag}(y)-I_n]$, where $I_n$ is the identity matrix of size $n$; the likelihood function of an LSBR model can then be written as
\begin{equation*}
	L\left(\beta; y \right) = \Lambda_n( B_y X \beta) \, .
\end{equation*}

\noindent
The next theorem shows that, if one restricts the CDF $\Lambda(\cdot)$ to be a scale mixture of Gaussian distributions, the $\psun$ class of priors is suitable for use in the Bayesian LSBR model.

\begin{restatable}{theorem}{pSUNconj}
	\label{pSUNconj}
	Consider a Bayesian LSBR model and assume that the prior for $\beta$ is 
	\begin{equation*}
		\beta \sim \psun_{p,m}( Q_{V}, \Theta, A, b, Q_W, \xi, \Omega ) \, .
	\end{equation*}
	Assume, in addition, that the link function $\Lambda(\cdot)$ has the following representation 
	\begin{equation*}
		\Lambda(x) = \int_{0}^{+\infty} \Phi \left(\frac{x}{\sqrt{v}}\right) dQ_{V^*}(v) \, .
	\end{equation*}
	Then, the posterior distribution of $\beta$ belongs to the $\psun$ family. More precisely,
	\begin{equation*}
		\beta \, \vert \, Y=y \sim \psun_{p,m+n} 
		\left(Q_V Q_{V^*}^n,
		\begin{bmatrix}
			\Theta & 0_{m \times n} \\
			0_{n \times m} & I_{n}
		\end{bmatrix}
		\begin{bmatrix}
			A & 0_{m \times p} \\
			0_{n \times p} & B_yX\mathrm{diag}^{\frac12}(\Omega)
		\end{bmatrix}, 
		\begin{bmatrix}
			b \\
			B_y X \xi
		\end{bmatrix},
		Q_W, \xi, \Omega
		\right) \, ,
	\end{equation*}
	where we have denoted, for $[x_1 , x_2]^\prime \in \mathbb{R}^{m+n}$,
	\begin{equation*}
		Q_{V} Q_{V^*}^n \left( [x_1 , x_2]^\prime
		\right) = Q_{V}(x_1) \prod_{i=1}^n Q_{V^*} (x_{2,i}) \, .
	\end{equation*}
\end{restatable}

\noindent
\textbf{Proof:}
See Supplementary Materials.

%\noindent
Theorem \ref{pSUNconj} shows that the $\psun$ family is a conjugate class for a large subset of LSBR, including probit and logit models. In this perspective, the previous theorem can be considered as a generalization of the results of \citet{durante2019}.
Furthermore, notice that a Gaussian distribution is a proper member of the $\psun$ family,
if one sets $m=0$ and $Q_W$ is the CDF of a point mass at $1_p$, a vector of $p$ $1$'s. 

From a computational point of view, we notice that, to perform the Gibbs algorithm to produce a posterior sample from the $\psun$ distribution, one must be able to sample from the full conditional distributions of $V$ and $W$, which reduce to $V \vert T$ and $W\vert Z$ respectively. Regarding the latter, this is relatively simple when the prior for $\beta$ either has an elliptical structure or has independent components. For example, the symmetric Generalized Hyperbolic \citep{barndorff1977} class of priors satisfies the elliptical constraint and corresponds to the $m=0$ case. Particular cases are the Gaussian and the Student-$t$ prior. In the former case, $W\vert Z$ is a constant equal to 1, in the latter case $W\vert Z$ has an Inverse Gamma distribution. When $m=1$, one obtains a new version of the skew Generalized Hyperbolic class.

Sampling from the full conditional of $V$ is generally difficult, depending on the link function $\Lambda(\cdot)$. However, computation is much simpler in the most common cases. For example, when the matrix ${\Theta}$ is diagonal and the components of $V$ are a priori independent, that is  $T_i \vert V_i \, , \ i = 1,2, \dots, m+n$ are mutually independent, it is possible to sample all $V_i \vert T_i \, , \ i = 1,2, \dots, n+m$ independently. This occurs, for example, when $m = 0$ or $m = 1$, which correspond to using a scale mixture of Gaussian or skew-normal priors, respectively.

The closed-form expression \eqref{eq:psunDDF} provides a way of computing the marginal probability of the observed $Y_i$'s for a given LSBR model. Nonetheless, its practical evaluation is cumbersome and it typically requires additional simulation. However, in statistical practice, the most used link functions are the probit and logit ones; while the former is considered in detail in \citet{durante2019}, in the next section we show how to deal with the logistic regression model. Specifically, we will show, assuming a multivariate Gaussian prior on the regression coefficients, that one can easily obtain, through a straightforward importance sampling algorithm, the normalizing constant of expression \eqref{eq:psunDDF} and, consequently, any posterior summary.

%%--------------------------------------------------------------------------------------------------
\section{Bayesian Logistic Regression} 
\label{sec:logit}
In this section we consider the popular logistic regression model; in Section \ref{subsec:gibbsLogis} we describe in detail a Gibbs sampler algorithm for producing a posterior sample from the distribution of the regression coefficients; in Section \ref{subsec:IS} we claim that computations are much easier to perform when the $\beta$ prior is Gaussian; under this assumption, the normalizing constant can be evaluated through an importance sampling algorithm: this opens the way to an easy calculation of posterior means and variances, and facilitates model selection procedures based on the Bayes factor.

As shown by \citet{andrews1974} and \citet{stefanski1991}, the logistic distribution admits a representation as a scale mixture of Gaussian densities, so the hypotheses on the link function of Theorem \ref{pSUNconj} are satisfied in the case of logistic regression. However, in terms of variance, the mixing density is four times the square of a Kolmogorov distribution \citep{kolmogorov1933, smirnov1939}; for the sake of clarity, we refer to this distribution as the \textit{logistic Kolmogorov}. In detail, let $\Qkol(\cdot)$ and $\qkol(\cdot)$ be the CDF and the density function of a Kolmogorov random variable; if $V_i = 4V^2_{0,i}$ and $V_{0,i} \sim \Qkol(\cdot)$ then $V_i$ has a logistic Kolmogorov distribution, and we denote it with $V_i \sim \LK(\cdot)$; the representation as a scale mixture of Gaussian densities of a standard logistic random variable can be rewritten in the following way:
\begin{equation}
	\label{eq:logisRepr}
	T_i \vert V_{i} \sim N(0, V_{i}) \, , \ V_{i} \sim \LK(\cdot) \Longrightarrow T_i \sim \Logis (0,1) \, ,
\end{equation}
and the density of $V_i$ is
\begin{equation*}
	f_{V_i}(v) = \lk(v) = \qkol \left( \frac{\sqrt{v}}{2} \right) \frac{1}{4 \sqrt{v}} \, ,
\end{equation*}
where the expression of $\qkol$ is reported in \citet{onorati2022} and reproduced in the Supplementary Materials. 
Theorem \ref{postLogisKolmo} lists some properties of the random variable
 $V_i \vert T_i = t  $, which will be useful from a computational perspective.

\begin{restatable}{theorem}{postLogisKolmo}
	\label{postLogisKolmo}
	Let $V_i \sim \LK(\cdot)$ and $T_i \vert V_i \sim N(0, V_i)$.
	Then:
	\begin{align*}
		\mathrm{\emph{A}.} &\hspace{1.9mm} M_{V_i}(u \vert T_i = t) = \mathrm{E}(e^{u V_i} \vert T_i = t) = e^{\abst} ( 1 + e^{-\abst} ) \sum_{k = 1}^{+\infty} (-1)^{k+1} k^2 \frac{ \exp \left( -\abst \sqrt{k^2 - 2u} \right) }{ \sqrt{k^2 - 2u} } \, , \\
		\mathrm{\emph{B}.} &\hspace{1.9mm} \mathrm{E}(V_i \vert T_i = t) = ( 1 + e^{-\abst} ) \left( \abst + (1 + e^{\abst}) \log(1 + e^{-\abst}) \right) \, .
	\end{align*}
\end{restatable}

\noindent
\textbf{Proof:}
See Supplementary Materials.

%\noindent
\citet{holmes2006} have already used the representation of a logistic density as a scale mixture of Gaussians in a data-augmentation Gibbs algorithm for a logistic regression. The approach described here and the one proposed by \citet{holmes2006} share some characteristics in the binary logistic case, although we introduce some improvements in terms of mixing and computational speed.

\subsection{Gibbs Algorithm for Logistic Regression.}
\label{subsec:gibbsLogis}
Theorem \ref{pSUNconj} guarantees that the resulting marginal posterior of $\beta$ still belongs to the $\psun$ family. 

Here we only describe how to sample from the conditional distributions of $V \vert T, \beta, W, Y$, which reduces to $V \vert T$: this task is particularly challenging in the logistic regression framework.

The first $m$ components of $V \vert T$ refer to the prior distribution and they are independent of the last $n$ ones, associated with the data. As a consequence, they are easy to manage for the most common priors. Here we focus on the last $n$ components of $V \vert T$, which involve the logistic Kolmogorov component. They are also mutually independent and we only need to discuss how to sample from $V_i \vert T_i \, , \, i = m+1, m+2, \dots, m+n$. To do that, we adopt a simple acceptance-rejection algorithm. The following theorem provides a bound between a suitable proposal density and the density of $V_i \vert T_i$.
\begin{restatable}{theorem}{accRej_plk}
	\label{accRej_plk}
	Let $V_i \sim \LK(\cdot)$, $T_i \vert V_i \sim N(0, V_i)$, $\widetilde{V} \sim \igamma(\alpha, \pi^2 / 2)$ and $\widetilde{T} \vert \widetilde{V} \sim N(0, \widetilde{V})$; set $\alpha > 3 / 2$.
	Then the ratio $f_{V_i}(v \vert T_i = t) / f_{\widetilde{V}}(v \vert \widetilde{T} = t)$ is bounded above by
	\begin{equation}
		\label{accRej_bound}
		M^\ast \frac{ f_{\widetilde{T}}(t) }{ f_{T_i}(t) } \, ,
	\end{equation}
	with $M^\ast = \max \Big( \max\limits_{0 < v \le v^\ast} \delta_1(v) \, , \, \max\limits_{v > v^\ast} \delta_2(v) \Big) $, $v^\star \in (1/2, 18 \pi^2 / 11)$,
	\begin{align*}
		\begin{bmatrix}
			\delta_1(v) \\
			\delta_2(v) 
		\end{bmatrix} &= \left\{
		\begin{alignedat}{2}
			& \frac{\sqrt{2 \pi^5} \Gamma(\alpha)} {(\pi^2/2)^\alpha} v^{\alpha - \frac 32} \quad &\mathrm{if}& \quad 0 < v \le v^\ast \\
			& \frac{\Gamma(\alpha)} {(\pi^2/2)^\alpha} v^{\alpha+1} \exp \left( \frac {\pi^2}{2 v} - \frac v 2 \right) \quad &\mathrm{if}& \quad v > v^\ast
		\end{alignedat}
		\right. \, , \\
		\arg \max_{0 < v \le v^\ast} \delta_1(v) &= v^\ast \, , \\
		\arg \max_{v > v^\ast} \delta_2(v) &= \left\{
	\begin{alignedat}{2}
		&1 + \alpha + \sqrt{(1+\alpha)^2 - \pi^2} \quad &\mathrm{if} \quad \alpha \ge \pi -1 \\
		&v^\ast &\mathrm{otherwise} \qquad \quad \ \,
	\end{alignedat}
		\right. \, .
	\end{align*}
\end{restatable}

\noindent
\textbf{Proof:}
See Supplementary Materials.

Notice that $f_{T_i}(\cdot)$ is the density of a standard logistic distribution; furthermore, it is well known that $\widetilde{V} \vert \widetilde{T} = t \sim \igamma \Big( \alpha + 1/2, (\pi^2 + t^2) / 2 \Big)$ and
\begin{equation*}
	f_{\widetilde{T}}(t) = \frac{\Gamma(\frac{2 \alpha + 1}{2})}{\Gamma(\alpha)\sqrt{ \pi^3}} \left( 1 + \frac{t^2}{\pi^2} \right)^{-\frac{2 \alpha + 1}{2}} \, .
\end{equation*}

\noindent
Expression \eqref{accRej_bound} provides the expected number of trials before accepting a value and it depends on the actual values of $|t|$ and 
$\alpha$. Efficient values for $\alpha$ are obtained by imposing
$\mathrm{E}( \widetilde V \vert \widetilde T = t) = \mathrm{E}( V_i \vert T_i = t)$.
This implies
\begin{equation*}
	\alpha = \frac12 \left( 1 + \frac{\pi^2 + t^2}{ \mathrm{E}( V_i \vert T_i = t) } \right) \, ,
\end{equation*}
and a simple closed-form expression of $\mathrm{E}( V_i \vert T_i = t)$ is provided by Theorem \ref{postLogisKolmo}. The density of a Kolmogorov r.v. has two alternative series expansions; while the former converges more quickly for small values of $v$, the latter is more suitable for large values of $v$; the threshold $v^\ast = 1.9834$ is the best compromise for optimizing the two rates of convergence.
The above proposal is very efficient since drawing values from the Inverse Gamma is cheap and the acceptance probability is always larger than 0.7 as long as $|t| \le 2750$. 
Even beyond that threshold, the algorithm performs relatively well; if $|t| = 10^6$, the acceptance probability is approximately $0.014$, but in this case, the algorithm can draw a sample of size $10^5$ in about $200$ milliseconds with a laptop with $4$ Gb of RAM and an i3-6100U as the CPU. 
As a final remark, we notice that the described method also provides a simpler way to sample from the Kolmogorov distribution \citep{devroye1984}. In fact, Theorem \ref{accRej_plk} gives an upper bound for the ratio between the logistic Kolmogorov and the Inverse Gamma densities. Then, one can sample $v\sim \LK(\cdot)$ with a standard accept-reject scheme and then take $\sqrt{v}/2$.

\subsection{Importance Sampling for Logistic Regression with a Gaussian Prior}
\label{subsec:IS}
In this section, we consider a particular, but important case where the prior distribution on $\beta$ is Gaussian, say $\beta \sim N_p(\xi, \Omega)$. In this situation, the $\psun$ is such that $W_i = 1 \, , \, i = 1,2, \dots, p$, say $Q_W(w) = \mathbb{I}_{ \{ w \ge 1_p \} }$; it follows that if one sets $V$ to some specific value, say $\Vfix$, then the posterior distribution reduces to a $\sun$ random variable. Our strategy is to adopt an ``importance sampling'' approach to completely avoid the use of an MCMC scheme. The importance density is based on a scale mixture of $\sun$ distributions; we refer to this random variable as a Unified Skew-$t$ ($\sut$) random variable since the mixing density is an Inverse Gamma. More in detail, if $\zeta_1 \sim \sut_{p,n}(\nu, \Theta, A, b, \xi, \Omega)$ then
\begin{align*}
	\zeta_1 &= \, \xi + \sqrt{S} \, \mathrm{diag}^\frac{1}{2}(\Omega) \zeta_0 \, , \\
	S \sim \igamma \left( \frac{\nu}{2}, \frac{\nu}{2} \right) &\indep \zeta_0 \sim \sun^\ast_{p,n}(\Theta, A, b, 0, \bar{\Omega}) \, .
\end{align*}
The following theorem gives the density of a $\sut$ random variable.
\begin{restatable}{theorem}{sutPDF}
	\label{sut}
The density of $\zeta_1 \sim \sut_{p,n}(\nu, \Theta, A, b, \xi, \Omega)$ is
	\begin{equation*}
		f_{\zeta_1}(x) = t_{\nu, \Omega}(x-\xi) \frac{
			T_{\nu+p, -b, \Theta} \left( \sqrt{ \frac{\nu+p}{\nu + (x-\xi)^\prime \, \Omega^{-1} \, (x-\xi)} } A \, \mathrm{diag}^{ -\frac{1}{2} }(\Omega) (x-\xi) \right)
		} {\Phi_{\Theta + A \, \OmegaBar A^\prime}(b)} \, ,
	\end{equation*}
 where $t_{\nu, \Omega}(\cdot)$ is the PDF of a Student-$t$ random variable with $\nu$ degrees of freedom and scale matrix $\Omega$ and $T_{\nu, b, \Theta}(\cdot)$ is the CDF of a random variable $R_1$ with the following stochastic representation:
	\begin{align} 
		\label{eq:mnct}
		\begin{split}
			&R_1 = \sqrt{S}(R_0 + b) \, , \\
			&S \sim \igamma \left( \frac{\nu}{2}, \frac{\nu}{2} \right) \indep R_0 \sim N_n(0, \Theta) \, .
		\end{split}
	\end{align}
\end{restatable}

\noindent
\textbf{Proof:} See Supplementary Materials.

\noindent
We notice that $R_1$ is a multivariate non-central Student-$t$ random vector. Furthermore, if $b$ is equal to the null vector, our construction of $\sut$ distribution coincides with the $\sut$ defined by \citet{jamalizadeh2010}: in this case, it belongs to the $\sue$ parametric family \citep{genton2010}.

In general, the computation of the CDF of a multivariate non-central Student-$t$ is cumbersome; however, it simplifies when $\Theta$ is a diagonal matrix. In the case of logistic regression with a Gaussian prior, $\Theta$ is the identity matrix. Then
\begin{equation*}
	T_{\nu+p, -b, I_n}(x) = \int_{0}^{+\infty} \prod_{i=1}^{n} \Phi \left( \frac{x_i} {\sqrt{s}} + b_{i} \right) f_{ \widehat{S} }(s) ds \, ,
\end{equation*}
where $f_{\widehat{S}}(\cdot)$ is the density function of an Inverse Gamma random variable with parameters $\big( (\nu+p)/2, (\nu+p)/2 \big)$. Let $u = F_{\widehat{S}}(s)$; we obtain
\begin{equation*}
	T_{\nu+p, -b, I_n}(x) = \int_{0}^{1} \prod_{i=1}^{n} \Phi \left( \frac{x_i} {\sqrt{ F^{-1}_{\widehat{S}}(u) }} + b_{i} \right) du \,,
\end{equation*}
and this expression suggests the use of a quasi-Monte Carlo algorithm, that is 
\begin{equation*}
	T_{\nu+p, -b, I_n}(x) \approx \frac{1}{k} \sum_{j=1}^k \prod_{i=1}^{n} \Phi \left( \frac{x_i} {\sqrt{ F^{-1}_{\widehat{S}}(u_j) }} + b_{i} \right) ,
\end{equation*}
where $u_j = {j}/{(k+1)}$ and we have set $k = 128$.

Therefore, in order to obtain a weighted sample from $\psun_{p,n}(Q_V, I_n, A, b, \mathbb{I}_{ \{ w \ge 1_p \} }, \xi, \Omega)$ we use $\sut_{p, n}(\nu_n, I_n, \diagInvOHf(\Vfix)A, \diagInvOHf(\Vfix)b, \xi^\star, \Omega)$  as the importance density. We set $\xi^\star = \beta_\mathrm{MaP} - \beta^\ast$; where $\beta_\mathrm{MaP}$ is the \textit{maximum a posteriori}, i.e. the posterior mode; and $\beta^\ast$ is the mode of $\sun^\ast_{p,n}(I_n, \diagInvOHf(\Vfix)A, \diagInvOHf(\Vfix)b, 0_p, \Omega)$, that is the mode of the importance density before translation and conditioned to $S = 1$. The following theorem guarantees that the above algorithm, which we denote $\psun$-IS,  produces a finite variance estimator for all finite sample sizes.

\begin{restatable}{theorem}{ISgood}
	\label{ISgood}
	In an LSBR model, if the posterior distribution is a $\psun_{p,n}(Q_V, I_n, A, b, \mathbb{I}_{ \{ w \ge 1_p \} }, \xi, \\ \Omega)$ and a $\sut_{p, n}(\nu_n, I_n, \diagInvOHf(\Vfix)A, \diagInvOHf(\Vfix)b, \xi^\star, \Omega)$ is used as the importance density, then
the importance weights are bounded above for every choice of $\xi^\star$, $\nu_n < +\infty$, $\Vfix$, and for every sample size $n$.
\end{restatable}

\noindent
\textbf{Proof:}
See Supplementary Materials.

\noindent
Notice that the computation of the two modes is relatively easy since both density functions are log-concave in the case of the logit model. We set $\nu_n = \max(100, n - p)$. Regarding the choice of $\Vfix$, a simple option satisfying the condition of the above theorem is to set $c = \pi^2 / 3$, which is the prior mean of each $V_i$. However, in our implementation we set $\Vfix = \mathrm{E}(V \vert \beta = \beta_{\mathrm{MaP}})$; in this way, based on our results, the algorithm is more efficient. Indeed, under the hypothesis of Theorem \ref{ISgood}, $\mathrm{E}(V \vert \beta = \beta_{\mathrm{MaP}})$ converges to $\mathrm{E}(V \vert Y)$ as $n$ goes to infinity; it is reasonable that setting $\Vfix$ equal to the posterior mean of $V$ is a better option than setting $\Vfix$ to the prior mean. Furthermore, the computation of $\mathrm{E}(V \vert \beta = \beta_\mathrm{MaP})$ is cheap using the expression of $\mathrm{E}(V \vert T)$ (see Theorem \ref{postLogisKolmo}). Indeed, set $a_i = A^\prime_i Z_\mathrm{MaP} + b_i$, where $A^\prime_i$ is the $i$-th row of $A$ and $Z_\mathrm{MaP} = \diagInvOH(\Omega)(\beta_\mathrm{MaP} - \xi)$; this implies that
\begin{equation}
	\label{postMeanVi}
	\mathrm{E}(V_i \vert \beta = \beta_\mathrm{MaP}) = \frac{1}{\mathrm{P}(T_i \le a_i)} \int_{-\infty}^{a_i} \mathrm{E}(V_i \vert T_i = t) f_{T_i}(t) dt \, .
\end{equation} 
The above expression can be easily evaluated using a quasi-Monte Carlo method, i.e.
\begin{equation*}
\mathrm{E}(V_i \vert \beta = \beta_\mathrm{MaP}) \approx \frac{1}{k} \sum_{j=1}^k \mathrm{E}(V_i \vert T = t_j) \, , \quad 
	t_j = \frac{u_j}{e^{-a_i}+1-u_j} \, ,
\end{equation*}
where, again, $u_j = {j}/{(k+1)}$. We used $k = 1024$.

%%--------------------------------------------------------------------------------------------------
\section{Examples}
\label{sec:exa}
 
This section is devoted to systematically exploring the finite sample performance of our proposal with simulated and real data.
Specifically, we have considered especially challenging situations, for example, when $p >> n$, or when the responses are highly unbalanced.
We have also tested the method by using several different prior specifications and link functions.

The performance of the new proposal has also been compared with that of the most popular and efficient alternatives available in the literature. In the case of logistic regression, we have included in the simulation the results obtained using the standard Polya-Gamma (PG) Gibbs sampler \citep{polson2013} and the more recent Ultimate Polya-Gamma (UPG) \citep{ultimate}.

In addition, the role of the prior is deemed important, especially when the sample size $n$ is small compared to $p$ and the estimators typically show large mean square errors. In this scenario, we have considered several alternative proposals, as discussed below. 

To be fair and to allow honest comparisons, all prior choices are vague, in the sense that they do not contain genuine prior information. In particular, all the $\psun$ priors considered share the following assumptions:  $m=0$, $\xi=0_p$, and a diagonal matrix for $\Omega$. This amounts to saying that all the priors are unimodal and symmetric around zero, although the corresponding $\psun$ posteriors will be skewed. 
In all cases, we have adopted weakly informative values for the hyperparameters in the spirit of \citet{gelman2008}. 

We have first implemented the probit model, which implies setting $V_1 = \cdots = V_n = 1$. 
Here we considered four different priors for the $\beta$ vector, namely, 
i) a Gaussian prior, which corresponds to setting $W_1=\cdots = W_p = 1$ \citep{durante2019}; 
ii) a multivariate (elliptical) Cauchy prior, obtained by setting $W_1= \dots = W_p = W^\ast$ with $W^\ast \sim \igamma(0.5, 0.5)$; 
iii) a multivariate Laplace with independent components; we will refer to it as \textit{Laplacit}, where $W_1,\dots W_p \overset{iid} \sim \textrm{ Exp }(1/2)$; and 
iv) a Dirichlet-Laplace prior \citep{bhattacharya2015}, with a discrete uniform prior on the Dirichlet parameter, with support $\{j/{300},\, j = 1, 2, \dots , 300\}$.
Finally, in all cases, the diagonal elements of $\Omega$ were assumed to be equal. Specifically, in the Gaussian case, its value was taken from \citet{durante2019} for the probit case; in the other cases, we have calibrated the scaling parameters to have roughly the same $95\%$ prior credible intervals as the Gaussian prior; all values are reported in Table \ref{priorPar}.

\noindent
With the logistic model, the $V_i$s are i.i.d. with a $\LK(\cdot)$ distribution. We have explored the performance of the same priors for $\beta$ as in the probit case. The diagonal components of $\Omega$ for the Gaussian prior are the ones used in the probit case multiplied by $\pi^2 / 3$, which is the expected value of a logistic Kolmogorov distribution; in all other cases, we have set the scaling parameters to have roughly the same prior credible intervals at level $95\%$ of the Gaussian case. Specific values are reported in Table \ref{priorPar}.
\begin{table}
    \caption{ Diagonal Elements of $\Omega$ for all Combinations of Models and Priors. }
	\label{priorPar}
    \begin{center}
        \begin{tabular}{|c|c|c|c|c|}
            \hline
            \textbf{Link} & \textbf{Gaussian} & \textbf{Cauchy} & \textbf{Laplacit} & \textbf{Dirichlet-Laplace} \\
            \hline
            Probit & 16 & 0.3807 & 6.8487 & 2.7655 \\
            \hline
            Logit  & 52.6379 & 1.2525 & 22.5314 & 9.0984 \\
            \hline
        \end{tabular}
    \end{center}
\end{table}
For all the simulation studies, we have reported the \textit{effective sample size per second}, which is the \textit{effective sample size} (ESS) divided by the computational time in seconds. More in detail, if $N$ is the number of simulations, in the importance sampling case, the ESS is computed as
\begin{equation*}
	\mathrm{ESS}_{IS} = \frac{1}{ \sum_{i=1}^N \widetilde{w}^2_i} \, ,
\end{equation*}
where the $\widetilde{w}^2_i$s are the normalized weights  
and, in MCMC cases, the ESS is defined as
\begin{equation}
	\label{eq:essMCMC}
	\mathrm{ESS}_{MCMC} = \frac{N}{1+2\sum_{i \geq 1} \rho(i)}, \,
\end{equation}
where $\rho(i)$ is the autocorrelation at lag $i$. To compute the quantity \eqref{eq:essMCMC} we have used the R package \texttt{sns}.
Finally, all of the computations are performed with sequential computing in an HPC with an Xeon E5-4617 as CPU; we have used $5$ Gb of RAM during the calculation.

\subsection{Simulation Study: Coverage Analysis} 
\label{subsec:IScov}
Here we illustrate the performance of the importance sampling algorithm for three different sample sizes, $n = 50, 100, 200$, and two values for the number of covariates, namely $p = 500, 1000$.
In all cases we have assumed a Gaussian prior and the logit link. We replicate the simulation $G = 10^4$ times for each pair $(n,p)$. 
The simulation scheme is described below. In the Supplementary Materials, we have reported the frequentist coverage for the intercept and the other parameters. Table \ref{IScoverage_ESSps} reports summaries of \textit{effective sample size per second} for all $(n,p)$. 
As noticed by a referee, the behavior of the ESS per second is not linear in $n$ and $p$. More precisely it worsens as $p$ increases for moderate sample sizes, say $n \approx 100$. On the other hand, for larger values of $n$, the ESS per second increases with the number of covariates. Such behavior is reminiscent of the double descent phenomenon observed in \citet{anceschi}, in the SUN case. See Supplementary Materials for other details. 
\begin{mdframed}
	\begin{tabbing}
		\qquad \enspace \small Set $X_{i,1} = 1 \, , \, i = 1, 2, \dots, n$. \\
		\qquad \enspace \small For $g = 1, 2, \dots,G$ \\
		\qquad \qquad $\bullet$ \small sample $X_{ij} \overset {\mathrm{iid}} \sim N(0,1) \, , \, i = 1, \dots, n \, ; \, j = 2,\dots, p$                              \\ 
		\qquad \qquad $\bullet$ \small centre and scale all columns of $X$, except the first, to have a standard \\
		\qquad \qquad \textcolor{white}{$\bullet$} \small deviation equal to $0.5$ \\
		\qquad \qquad $\bullet$ \small sample $\beta^{\dagger,g}$ from its prior \\
		\qquad \qquad $\bullet$ \small sample $Y_i \overset {\mathrm{ind}} \sim Be \Big( \Lambda(X^\prime_i \, \beta^{\dagger,g} ) \Big) \, , \, i = 1,2, \dots, n$  \\
		\qquad \qquad $\bullet$ \small draw $N = 10^4$ values from the posterior distribution of $\beta$ through importance \\ \textcolor{white}{\qquad \qquad $\bullet$} sampling \\
		\qquad \qquad $\bullet$ \small compute the empirical quantiles of level $\gamma \in \{5/100 \times j, j = 1,2,\dots,19\}$ \\
		\qquad \enspace \small $\Longrightarrow$ evaluate the results in terms of frequentist coverage, \\
		\textcolor{white}{\qquad \enspace \small $\Longrightarrow$} {effective sample size} and computing time
	\end{tabbing}
\end{mdframed}

\centerline{\textbf{Simulation Scheme for Coverage Analysis}}

\begin{table}[ht]
	\caption{ Coverage Analysis of $\psun$-IS: ESS per second in the logistic regression. Different Combinations of Sample Size $n$ and Number of Parameters $p$. Number of simulations $N = 10^4$. }
	\label{IScoverage_ESSps}
	\begin{center}
		\begin{tabular}{|c c|c c c c c c|}
			\hline
			\textbf{n} & \textbf{p} & \textbf{Min} & \textbf{1st Qu.} & \textbf{Median} & \textbf{Mean} & \textbf{3rd Qu.} & \textbf{Max} \\
			\hline
			50 & 500 & 322.15 & 384.18 & 391.35 & 390.05 & 397.15 & 445.35 \\ 
			50 & 1000 & 194.64 & 220.31 & 223.74 & 223.34 & 226.70 & 252.28 \\ 
			100 & 500 & 120.79 & 153.03 & 157.56 & 156.92 & 161.51 & 182.28 \\ 
			100 & 1000 & 97.01 & 113.77 & 116.28 & 115.90 & 118.39 & 132.86 \\ 
			200 & 500 & 2.47 & 4.63 & 5.22 & 5.29 & 5.86 & 10.94 \\ 
			200 & 1000 & 16.05 & 24.31 & 26.01 & 25.96 & 27.65 & 36.80 \\ 
			\hline
		\end{tabular}
	\end{center}
\end{table}
\subsection{Simulation Study: Polya-Gamma vs \texorpdfstring{$\boldsymbol{\psun}$}{\psun} with Small Sample Size}
\label{subsec:PGvs_pSUN}
In this subsection, we evaluate the performances of different algorithms in the case of a logit model under Gaussian, Laplacit, and Dirichlet-Laplace priors. The methods under comparison are the Polya-Gamma Gibbs sampler, the Ultimate Polya-Gamma sampler, the $\psun$-Gibbs, and the $\psun$-IS, the last one only in the case of a Gaussian prior. We have set $n = 50$ and $p = 500$. The number of posterior draws was $N = 10^4$, but with the Dirichlet-Laplace prior we have used $N=10^5$.  The entire simulation was replicated $G = 2400$ times.
The simulation scheme is similar to that described in \S~\ref{subsec:IScov}, with the only difference that the true $\beta$ are kept fixed according to the following values:
{\small
\begin{align*}
\beta^\dagger_1 =\Lambda^{-1}(0.50)\, = 0 \,,     \;\;\,\quad\qquad           &\qquad \beta^\dagger_2,   \dots, \beta^\dagger_{26} =\Lambda^{-1}(0.05)=-2.9444\,, \\
\beta^\dagger_{27} , \dots, \beta^\dagger_{76} = \Lambda^{-1}(0.10)\,=-2.1972, &\qquad \beta^\dagger_{77}, \dots, \beta^\dagger_{126}=\Lambda^{-1}(0.20)\,=-1.3863, \\ 
\beta^\dagger_{127}, \dots, \beta^\dagger_{176} = \Lambda^{-1}(0.30)\,=-0.8473, &\qquad \beta^\dagger_{177}, \dots, \beta^\dagger_{226}=\Lambda^{-1}(0.40)\,=-0.4055, \\
\beta^\dagger_{227}, \dots, \beta^\dagger_{276} = \Lambda^{-1}(0.60)\,=0.4055, &\qquad \beta^\dagger_{277}, \dots, \beta^\dagger_{326}=\Lambda^{-1}(0.70)\,=0.8473, \\
\beta^\dagger_{327}, \dots, \beta^\dagger_{376} = \Lambda^{-1}(0.80)\,=1.3863, &\qquad \beta^\dagger_{377}, \dots, \beta^\dagger_{426}=\Lambda^{-1}(0.90)\,=2.1972, \\
\beta^\dagger_{427}, \dots, \beta^\dagger_{451} = \Lambda^{-1}(0.95)\,=2.9444, &\qquad \beta^\dagger_{452}, \dots, \beta^\dagger_{500}=\Lambda^{-1}(0.50)\,= 0, %\
\end{align*}
}

\noindent
where $\Lambda(\cdot)$ is the CDF of the standard logistic distribution.

\begin{center}
	\begin{figure}
		\caption{ Polya-Gamma vs $\psun$ with small sample size: Mean of ACF for the Intercept, $\psun$-Gibbs (continuous line) vs PG (dashed line) vs UPG (dotted line). Left: Gaussian Prior. Center: Laplacit Prior. Right: Dirichlet-Laplace Prior. }
		\centerline{\includegraphics[scale=1]{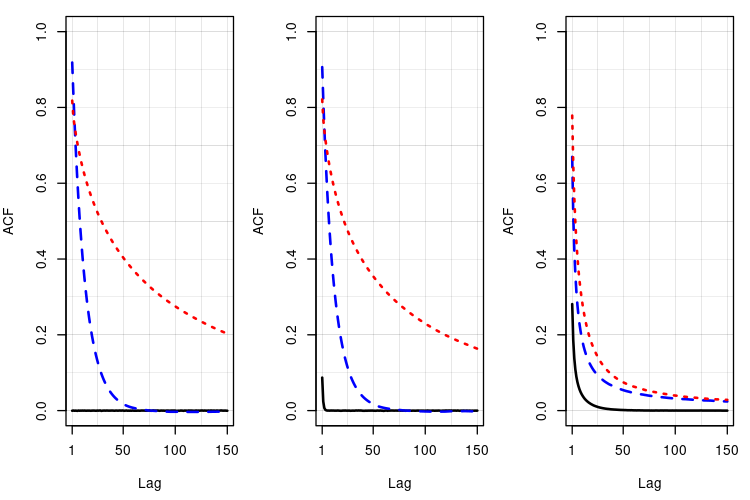}}
		\label{PGvspSUN_interceptACF}
	\end{figure}
\end{center}

\noindent
Figure \ref{PGvspSUN_interceptACF} reports the average of the ACF for the intercept using Gaussian (a), Laplacit (b), and Dirichlet-Laplace (c) priors. The superior mixing of $\psun$-Gibbs is apparent. Nonetheless, for the other parameters, the difference is minor; however, when a Dirichlet-Laplace prior is adopted, differences between Polya-Gamma and $\psun$ algorithms are more evident for all $\beta_i$s.
In all cases, the $\psun$-Gibbs method shows better mixing; in particular, posterior draws using a  Gaussian prior are essentially uncorrelated. With the Dirichlet-Laplace prior, the chains obtained using Polya-Gamma samplers do not seem to converge to the posterior distribution; even after $100\,000$ simulations, all the empirical means of the posterior samples
are roughly $0$. We suspect that this occurs because, in these cases, the MCMC chains can hardly get out from the region near the origin. 
In other terms, for finite samples, under a Dirichlet-Laplace prior, one has
\begin{equation*}
	\lim_{ \lVert \beta \rVert \rightarrow 0} f_\beta(\beta \vert Y) = + \infty \, .
\end{equation*}
In the Supplementary Materials, we provide a theoretical motivation for the above empirical findings. The better mixing of the $\psun$ algorithm can be related to the interpretation of the logit regression as a mixture of unidentifiable probit models. This implies that, conditionally on the logistic Kolmogorov random variables, the full conditional distribution of $\beta$ is a $\sun$ density and it retains a selection mechanism similar to that of the $\psun$. On the other hand, conditionally on the Polya-Gamma random variables as in \cite{polson2013}, the full conditional of $\beta$ is Gaussian, and it is not able to adequately represent the posterior dependence structure.  
A similar argument, with minor modifications, can be used when the prior has a representation in terms of a scale mixture of Gaussian densities.

Despite its better convergence rate, the $\psun$-Gibbs becomes inefficient in terms of computational time as the sample size increases. Indeed, it relies on the algorithm of \citet{botev2017} to sample from truncated Gaussian distributions. This algorithm remains fast and efficient as long as the normalizing constant of the truncated Gaussian distribution is not too small. In general, Botev's algorithm performs well for small sample sizes, say $n \le 100$; for larger values of $n$, it is only reliable when the components of the truncated Gaussian are weakly correlated, and this is likely to occur when the ratio $p / n$ is large. 
The same issue arises in the algorithm described in \citet{durante2019}. On the other hand, our Gibbs sampler and the exact i.i.d. random generator of \cite{durante2019} cover the case of large $p / n$ for logit and probit models respectively, and precisely in this scenario the most common MCMC methods show poor performances.

In the Supplementary Materials, we have reported the means of the \textit{effective sample size per second} under the different algorithms. The $\psun$ methods clearly show a better performance; indeed, the $\psun$-Gibbs is at least $2$ times faster than the Polya-Gamma methods in terms of ESS per second, and in many cases, the difference is much larger, specifically in the case of the intercept and the Dirichlet-Laplace prior. However, when applicable, the $\psun$-IS largely outperforms all competitors in terms of ESS per second; this occurs because methods based on the Gibbs sampler must compute a matrix inversion at each iteration, while the importance sampling does this task only once at the beginning of the algorithm.

Table \ref{PGvs_pSUN_MSE} reports the mean squared errors (MSE) for the different groups of $\beta$ and different priors. 
The Dirichlet-Laplace prior typically shows the lowest MSEs. It is worth noting that, in the case of Gaussian prior, the $\psun$-Gibbs and $\psun$-IS produce the same MSEs, that is, using the importance sampling approach does not imply a degradation in terms of the quality of inference but computational times are much different: $\psun$-Gibbs requires in mean roughly $3$ mins and $45$ secs, $\psun$-IS is completed in mean on only $12$ secs, as reported in Table \ref{PGvs_pSUN_times}.

Nevertheless, there is a trade-off between computational and inferential efficiency; the Gaussian prior typically shows a larger mean square error but it is the most efficient one in terms of \textit{effective sample size per second}; on the contrary, the Dirichlet-Laplace typically shows a smaller mean square error but it is the least efficient in terms of \textit{effective sample size per second}. See Supplementary Materials for more details.

\begin{table}[ht]
	\caption{ Polya-Gamma vs $\psun$ with small sample size: MSE for the Different Groups of $\beta$ and Priors with $\psun$-Gibbs and pSUN-IS Algorithms. Number of simulations $N = 10^4$ for Gaussian and Laplacit and $N = 10^5$ for Dirichlet-Laplace. }
	\label{PGvs_pSUN_MSE}
	\begin{center}
		\begin{tabular}{ | c | c c c | c | }
			\hline
			\multirow{2}{*}{} & \multicolumn{3}{c|}{\textbf{$\bm \psun$-Gibbs}} & \textbf{$\bm \psun$-IS} \\
			& \textbf{Gaussian} & \textbf{Laplacit} & \textbf{Dirichlet-Laplace} & \textbf{Gaussian} \\
			\hline
            Intercept & 13.38 & 8.34 & 0.41 & 13.21 \\ 
            $\beta = \Lambda^{-1}(0.05)$ & 7.43 & 7.40 & 8.82 & 7.42 \\ 
            $\beta = \Lambda^{-1}(0.10)$ & 5.47 & 5.24 & 5.25 & 5.45 \\ 
            $\beta = \Lambda^{-1}(0.20)$ & 4.05 & 3.65 & 2.69 & 4.02 \\ 
            $\beta = \Lambda^{-1}(0.30)$ & 3.43 & 2.99 & 1.55 & 3.40 \\ 
            $\beta = \Lambda^{-1}(0.40)$ & 3.15 & 2.71 & 1.07 & 3.12 \\ 
            $\beta = \Lambda^{-1}(0.50)$ & 3.08 & 2.59 & 0.92 & 3.05 \\ 
            $\beta = \Lambda^{-1}(0.60)$ & 3.12 & 2.70 & 1.12 & 3.09 \\ 
            $\beta = \Lambda^{-1}(0.70)$ & 3.47 & 3.00 & 1.65 & 3.43 \\ 
            $\beta = \Lambda^{-1}(0.80)$ & 4.03 & 3.66 & 2.66 & 4.00 \\ 
            $\beta = \Lambda^{-1}(0.90)$ & 5.48 & 5.25 & 5.24 & 5.46 \\ 
            $\beta = \Lambda^{-1}(0.95)$ & 7.44 & 7.34 & 8.52 & 7.43 \\ 
			\hline
		\end{tabular}
	\end{center}
\end{table}
\begin{table}[ht]
	\caption{ Polya-Gamma vs $\psun$ with small sample size: Summaries of Computational Time (Seconds) with Different Algorithms and Priors. Number of simulations $N = 10^4$ for Gaussian and Laplacit and $N = 10^5$ for Dirichlet-Laplace. }
	\label{PGvs_pSUN_times}
	\begin{center}
	\scalebox{0.90}{
			\begin{tabular}{| c c | c c c c c c |}
				\hline
				\textbf{Algorithm} & \textbf{Prior} & \textbf{Min} & \textbf{1st Qu.} & \textbf{Median} & \textbf{Mean} & \textbf{3rd Qu.} & \textbf{Max} \\
				\hline
				UPG & Gaussian & 224.73 & 240.96 & 246.14 & 245.58 & 250.85 & 283.86 \\ 
				PG & Gaussian & 247.59 & 267.87 & 276.49 & 279.95 & 291.13 & 327.36 \\ 
				$\psun$-Gibbs & Gaussian & 198.15 & 216.27 & 223.64 & 226.72 & 236.17 & 271.68 \\ 
				$\psun$-IS & Gaussian & 11.25 & 11.70 & 11.97 & 12.28 & 12.80 & 20.77 \\ 
				UPG & Laplacit & 332.64 & 350.44 & 354.60 & 356.18 & 361.86 & 412.63 \\ 
				PG & Laplacit & 336.77 & 356.29 & 361.33 & 367.21 & 373.70 & 435.50 \\ 
				$\psun$-Gibbs & Laplacit & 310.35 & 328.38 & 335.82 & 340.85 & 349.47 & 401.18 \\ 
				UPG & Dirichlet-Laplace & 5934.51 & 8143.24 & 8550.91 & 8503.14 & 8916.34 & 11335.37 \\ 
				PG & Dirichlet-Laplace & 4403.25 & 5062.45 & 5407.70 & 5434.88 & 5719.30 & 6943.32 \\ 
				$\psun$-Gibbs & Dirichlet-Laplace & 3317.94 & 3682.11 & 3811.16 & 3882.04 & 4070.48 & 4770.39 \\ 
				\hline
			\end{tabular}
			}
		\end{center}
\end{table}

\subsection{Simulation Study: a Logit Model with Unbalanced Data} 
\label{subsec:unbalanced}

Here we consider the case of unbalanced data, that is when $\sum_{i=1}^n Y_i/n$ is either close to $0$ or $1$. In this scenario, using a logistic model, the Polya-Gamma Gibbs sampler might perform unsatisfactorily, as noticed in \citet{johndrow2019}. We show, at least empirically, that the same effect does not occur when using the $\psun$ Gibbs sampler.

\begin{center}
	\begin{figure}
		\caption{ Logit model with Unbalanced Data: ACF Comparison, $\psun$-Gibbs (continuous line) vs PG (dashed line) vs UPG (dotted line). Left: $n = 50$. Center: $n = 200$. Right: $n = 1000$. }
		\centerline{\includegraphics[scale=1]{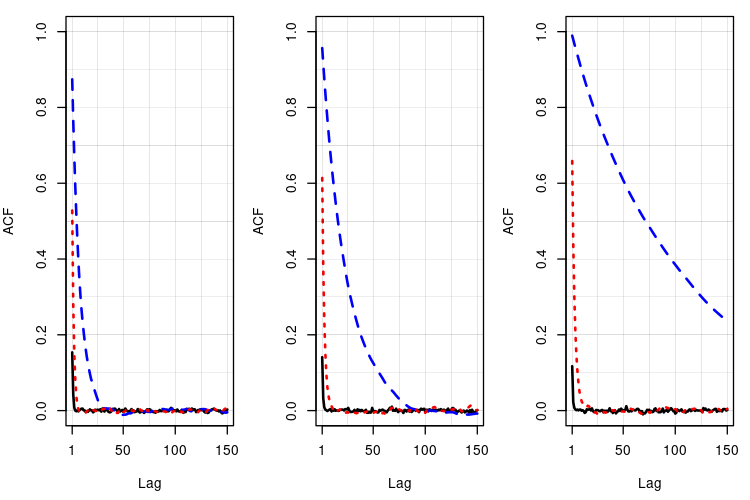}}
		\label{logitUnb_ACF}
	\end{figure}
\end{center}

We have replicated the simulation study described in \citet{johndrow2019}, by adopting a Gaussian prior, no covariates, and assuming $\sum_{i=1}^n Y_i = 1$; the experiment was repeated for sample sizes $n = 50, 200, 1000$. 
We have also run the Polya-Gamma and Ultimate Polya-Gamma samplers for comparison. We have run a chain of $N=10^5$ posterior draws in all cases.
Figure \ref{logitUnb_ACF} reports the ACFs.
The $\psun$-Gibbs shows a better mixing; the ACFs are insensitive to the sample size and draws are typically uncorrelated after roughly $4$ lags. On the other hand, the convergence issues detected by \citet{johndrow2019} arise for the Polya-Gamma approach; the Ultimate Polya-Gamma performs better in terms of ESS per second as shown in Table \ref{logitUnb_tab}. 
Notice that the $\psun$-Gibbs is mainly designed for small sample sizes and a large number of covariates; this simulation study considers precisely the opposite scenario; in fact, when $n = 1000$, the $\psun$-Gibbs is highly inefficient and it takes approximately $3.6$ seconds for each draw.
\begin{table}[ht]
	\caption{ Logit Model with Unbalanced Data: \textit{effective sample size per second} (ESSps), \textit{effective sample size} (ESS), and Computational Time in seconds for the Different Sample Sizes ($n$). Number of simulations $N = 10^5$. }
	\label{logitUnb_tab}
	\begin{center}
		\begin{tabular}{ | c | c | c c c |}
			\hline
			& \textbf{n} & \textbf{UPG} & \textbf{PG} & \textbf{$\bm \psun$-Gibbs} \\
			\hline
			ESSps & 50 & 126.03 & 51.69 & 70.82 \\ 
			ESSps & 200 & 25.85 & 4.52 & 9.50 \\
			ESSps & 1000 & 4.36 & 0.21 & 0.21 \\
   \hdashline
  		ESS & 50 & 30891.02 & 6616.94 & 70518.21 \\ 
			ESS & 200 & 23351.56 & 2179.75 & 73486.13 \\ 
			ESS & 1000 & 19457.58 & 529.75 & 77673.21 \\ 
   \hdashline
  		Time & 50 & 245.11 & 128.01 & 995.74 \\  
  		Time & 200 & 903.22 & 482.40 & 7738.85 \\   
  		Time & 1000 & 4458.04 & 2495.78 & 361322.89 \\ 
			\hline
		\end{tabular}
	\end{center}
\end{table}

\subsection{Simulation Study: a Probit Model with Sparse Parameters}
\label{subsec:sparse}

In this section, we compare the performance of Gaussian, Laplacit, and Dirichlet-Laplace priors in a probit scenario, when the parameters are sparse, let's say only a few values of $\beta^\dagger_i$s are not $0$. 
When $p/n$ is large, Botev's (\citeyear{botev2017}) algorithm is typically efficient; nevertheless, the impact of the prior can be significant. It is then important to adequately calibrate the hyper-parameters of the Gaussian prior; this is a difficult task without significant genuine prior information. A scale mixture of Gaussian densities allows us to learn the scale parameters from the data; furthermore, the use of a shrinkage prior might be useful to obtain robust results \citep{tibshirani1996}.
We fixed the sample size at $n = 50$, the number of parameters at $p = 500$, and the posterior draws at $N = 10^4$, but with the Dirichlet-Laplace prior we have used $N=10^5$. We then replicate the entire simulation $G = 2400$ times, keeping the true parameter vector $\beta^\dagger$ fixed. More in detail, we set
{\small
\begin{align*}
\beta^\dagger_1=\Phi^{-1}(0.50)= 0 \,,         \;\;\,\quad\qquad &                    \qquad \beta^\dagger_2,     \dots, \beta^\dagger_{400} = \Phi^{-1}(0.50)=0\,, \\
\beta^\dagger_{401}, \dots, \beta^\dagger_{410} = \Phi^{-1}(0.05)=-1.6449\,, &\qquad \beta^\dagger_{411}, \dots, \beta^\dagger_{420} = \Phi^{-1}(0.10)=-1.2816\,, \\ 
\beta^\dagger_{421}, \dots, \beta^\dagger_{430} = \Phi^{-1}(0.20)=-0.8416\,, &\qquad \beta^\dagger_{431}, \dots, \beta^\dagger_{440} = \Phi^{-1}(0.30)=-0.5244\,, \\
\beta^\dagger_{441}, \dots, \beta^\dagger_{450} = \Phi^{-1}(0.40)=-0.2533\,, &\qquad \beta^\dagger_{451}, \dots, \beta^\dagger_{326} = \Phi^{-1}(0.60)=0.2533\,, \\
\beta^\dagger_{461}, \dots, \beta^\dagger_{470} = \Phi^{-1}(0.70)=0.5244\,, &\qquad \beta^\dagger_{471}, \dots, \beta^\dagger_{426} = \Phi^{-1}(0.80)=0.8416\,, \\
\beta^\dagger_{481}, \dots, \beta^\dagger_{490} = \Phi^{-1}(0.90)=1.2816\,, &\qquad \beta^\dagger_{491}, \dots, \beta^\dagger_{500} = \Phi^{-1}(0.95)=1.6449\,, %%
\end{align*}
}

\noindent
where $\Phi(\cdot)$ is the CDF of the standard Gaussian distribution. The simulation scheme is similar to that described in \S~\ref{subsec:PGvs_pSUN}. Table \ref{probitSparse_MSE} reports the empirical MSE for all groups of $\beta$ and different priors, Table \ref{probitSparse_ESSps} reports the \textit{effective sample sizes per second}. As anticipated, the Dirichlet-Laplace generally performs better in terms of MSE, at least for $\vert \beta^{\dagger}\vert \leq \Phi^{-1}(0.8)$. See Supplementary Materials for more details.

\begin{table}[ht]
	\caption{ Probit Model with Sparse Parameters: MSE for the Different Groups of $\beta$ and Priors. Number of simulations $N = 10^4$ for Gaussian and Laplacit and $N = 10^5$ for Dirichlet-Laplace. }
	\label{probitSparse_MSE}
	\begin{center}
		\begin{tabular}{| c | c c c |}
			\hline
			& \textbf{Gaussian} & \textbf{Laplacit} & \textbf{Dirichlet-Laplace} \\ 
			\hline
  		Intercept & 4.65 & 3.34 & 0.18 \\ 
  		$\Phi^{-1}(0.05)$ & 1.56 & 1.61 & 2.91 \\ 
  		$\Phi^{-1}(0.10)$ & 1.31 & 1.30 & 1.98 \\ 
  		$\Phi^{-1}(0.20)$ & 1.10 & 1.01 & 1.00 \\ 
  		$\Phi^{-1}(0.30)$ & 1.01 & 0.88 & 0.54 \\ 
  		$\Phi^{-1}(0.40)$ & 0.95 & 0.80 & 0.33 \\ 
  		$\Phi^{-1}(0.50)$ & 0.94 & 0.80 & 0.28 \\ 
  		$\Phi^{-1}(0.60)$ & 0.94 & 0.82 & 0.34 \\ 
  		$\Phi^{-1}(0.70)$ & 0.99 & 0.88 & 0.54 \\ 
  		$\Phi^{-1}(0.80)$ & 1.11 & 1.03 & 1.01 \\ 
  		$\Phi^{-1}(0.90)$ & 1.32 & 1.28 & 1.79 \\ 
  		$\Phi^{-1}(0.95)$ & 1.58 & 1.62 & 2.85 \\ 
			\hline
		\end{tabular}
	\end{center}
\end{table}
\begin{table}[ht]
	\caption{ Probit Model with Sparse Parameters: Mean of ESS per second for the Different Groups of $\beta$ and Priors. Number of simulations $N = 10^4$ for Gaussian and Laplacit and $N = 10^5$ for Dirichlet-Laplace. }
	\label{probitSparse_ESSps}
	\begin{center}
		\begin{tabular}{| c | c c c |}
			\hline
			& \textbf{Gaussian} & \textbf{Laplacit} & \textbf{Dirichlet-Laplace} \\ 
			\hline
			Intercept & 5463.10 & 25.50 & 10.78 \\ 
			$\Phi^{-1}(0.05)$ & 5463.10 & 26.55 & 7.07 \\ 
			$\Phi^{-1}(0.10)$ & 5463.10 & 27.12 & 7.62 \\ 
			$\Phi^{-1}(0.20)$ & 5463.10 & 27.66 & 8.17 \\ 
			$\Phi^{-1}(0.30)$ & 5463.10 & 27.90 & 8.40 \\ 
			$\Phi^{-1}(0.40)$ & 5463.10 & 28.04 & 8.49 \\ 
			$\Phi^{-1}(0.50)$ & 5463.10 & 28.05 & 8.54 \\ 
			$\Phi^{-1}(0.60)$ & 5463.10 & 28.02 & 8.47 \\ 
			$\Phi^{-1}(0.70)$ & 5463.10 & 27.90 & 8.40 \\ 
			$\Phi^{-1}(0.80)$ & 5463.10 & 27.59 & 8.13 \\ 
			$\Phi^{-1}(0.90)$ & 5463.10 & 27.12 & 7.59 \\ 
			$\Phi^{-1}(0.95)$ & 5463.10 & 26.50 & 7.11 \\ 
			\hline
		\end{tabular}
	\end{center}
\end{table}
\subsection{ A Real Dataset: Parameters Estimation for \textit{Cancer SAGE}}
\label{subsec:CancerSAGE}

\begin{center}
	\begin{figure}
		\caption{ Cancer SAGE Example, Posterior Means of the 516 $\beta$ Coefficients Plus The Intercept $\beta_1$ Using the $\psun$-Gibbs Algorithm. Top: Logit. Bottom: Probit. Left: Gaussian Prior. Center: Laplacit Prior. Right: Dirichlet-Laplace Prior. }
		\label{CancerSAGE_est}
		\centerline{\includegraphics[scale=1]{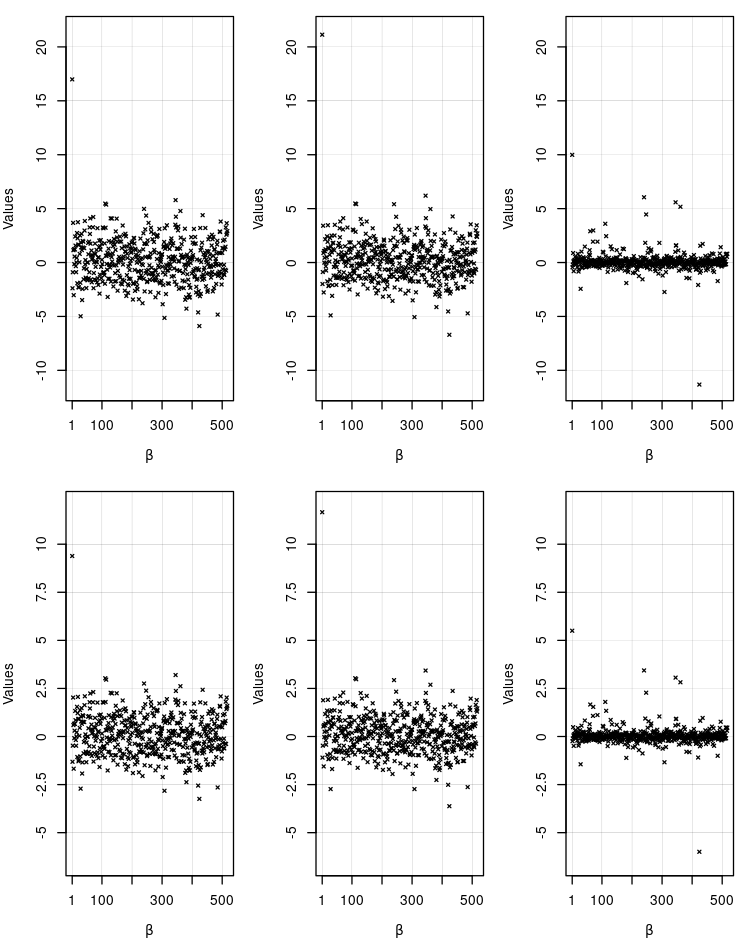}}
	\end{figure}
\end{center}

The ``Cancer SAGE" dataset has been discussed in \citet{durante2019} and represents a situation where the number of parameters is larger than the sample size. It is available online\footnote{\scriptsize \url{https://webusers.i3s.unice.fr/~pasquier/web/?Research_Activities___Dataset_Downloads___Cancer_SAGE}} and consists of the gene expressions of $n = 74$ normal and cancerous biological tissues at $516$ different tags. It is of interest to quantify the effects of gene expressions on the probability of a cancerous tissue and to predict the status of new tissues as a function of the gene expression. We have standardized the gene expressions to have a zero mean and a standard deviation equal to $0.5$. We have considered all the alternative priors defined before, including the Cauchy prior, combined with the logistic and probit regression models; for all cases, we obtain $N = 10^5$ posterior draws. In addition, in the logit case, we have compared the performances of our approach with both Polya-Gamma algorithms.
Figure \ref{CancerSAGE_est} reports the posterior means of all the $\beta$ coefficients using different priors, both in the logit and probit scenarios; we have excluded the results with the Cauchy prior because the posterior mean does not exist in that case. One can immediately notice that, with $n<<p$, the final results are typically driven by the prior assumptions; most of the variability among estimates can be justified by the differences in tail thickness and by the choices of the $\Omega$ values. However, based on the results of Sections \ref{subsec:PGvs_pSUN} and \ref{subsec:sparse}, we claim that the most robust estimates are those obtained using the Dirichlet-Laplace prior.
In the logit case, while the final inferences are practically identical using $\psun$ or Polya-Gamma methods (except for the Dirichlet-Laplace prior, where the Polya-Gamma algorithms simply do not converge), the computing \textit{effective sample sizes per second} are quite different, as shown in Table \ref{CancerSAGE_logitESSps}. The $\psun$ Gibbs sampler is at least $2$ times faster in terms of \textit{effective sample size per second} and, in some cases, the difference is much larger, i.e. for the intercept, where the Polya-Gamma algorithm shows the worst performance. 

Furthermore, in the case of the Gaussian prior, the importance sampling technique can greatly improve the computational times, compared with MCMC methods. Table \ref{CancerSAGE_probitESSps} reports the \textit{effective sample sizes per second} in the probit case.

Regarding the use of the Cauchy prior, we have implemented it both in the probit and the
logit models; for the latter, we have used the $\psun$ approach and both Polya-Gamma Gibbs
samplers. In this case, differences in performance among alternative methods are minor; in the
Supplementary Materials we show the behavior of the posterior medians of the coefficients since
posterior means do not exist.

\begin{table}[ht]
	\caption{Cancer SAGE Example, Logit Model: Summaries of ESS per second using different algorithms and priors. Number of simulations $N = 10^5$.} 
	\label{CancerSAGE_logitESSps}
	\begin{center}
	\scalebox{0.95}{
		\begin{tabular}{| c c | c c c c c c |}
			\hline
			\textbf{Algorithm} & \textbf{Prior} & \textbf{Min} & \textbf{1st Qu.} & \textbf{Median} &  \textbf{Mean} & \textbf{3rd Qu.} & \textbf{Max} \\
			\hline
			UPG & Gaussian & 0.21 & 1.75 & 2.55 & 3.13 & 3.90 & 16.55 \\ 
  		PG & Gaussian & 1.12 & 12.46 & 14.40 & 14.93 & 16.83 & 30.12 \\ 
  		$\psun$-Gibbs & Gaussian & 29.67 & 30.91 & 30.91 & 30.90 & 30.91 & 33.26 \\ 
  		$\psun$-IS & Gaussian & 209.81 & 209.81 & 209.81 & 209.81 & 209.81 & 209.81 \\
			\hdashline 
  		UPG & Laplacit & 0.08 & 1.35 & 1.87 & 2.26 & 2.76 & 9.96 \\ 
  		PG & Laplacit & 0.43 & 8.21 & 9.62 & 9.92 & 11.42 & 20.40 \\ 
  		$\psun$-Gibbs & Laplacit & 9.36 & 18.83 & 20.83 & 20.05 & 22.04 & 23.28 \\
			\hdashline 
  		UPG & Dirichlet-Laplace & 0.03 & 0.19 & 0.38 & 0.47 & 0.69 & 2.13 \\ 
  		PG & Dirichlet-Laplace & 0.06 & 0.46 & 0.99 & 1.24 & 1.86 & 5.57 \\ 
 		  $\psun$-Gibbs & Dirichlet-Laplace & 0.08 & 0.67 & 1.13 & 1.21 & 1.71 & 2.70 \\ 
			\hline
		\end{tabular}
	}
	\end{center}
\end{table}
\begin{table}[ht]
	\caption{Cancer SAGE Example, Probit Model: Summaries of ESS per second using different priors Number of simulations $N = 10^5$.}
	\label{CancerSAGE_probitESSps}
	\begin{center}
		\begin{tabular}{| c | c c c c c c |}
			\hline
			\textbf{Prior} & \textbf{Min} & \textbf{1st Qu.} & \textbf{Median} & \textbf{Mean} & \textbf{3rd Qu.} & \textbf{Max} \\
			\hline
			Gaussian & 1844.57 & 1844.57 & 1844.57 & 1844.57 & 1844.57 & 1844.57 \\ 
			Laplacit & 9.52 & 18.83 & 20.85 & 20.04 & 22.00 & 23.13 \\ 
			Dirichlet-Laplace & 0.08 & 0.69 & 1.14 & 1.21 & 1.71 & 2.75 \\ 
			\hline
		\end{tabular}
	\end{center}
\end{table}
\subsection{A Real Dataset: Model and Variable Selection for \textit{Lung Cancer}}
\label{subsec:LungCancer}

The lung cancer dataset has already been used by \citet{hong1991} and it is available online\footnote{\url{http://archive.ics.uci.edu/dataset/62/lung+cancer}}.
It contains $57$ attributes relating to $32$ patients with three types of lung cancer. The first attribute specifies the kind of tumor, while the others are used as predictors; the Authors do not provide much additional information for those variables. We have removed attributes $5$ and $39$ since they contained NAs; we have dichotomized the first attribute and set $Y_i = 1$ when it is equal to $1$ for the $i$-th patient; we have also included an intercept term: the model is then based on $n = 32$ observations and $p = 55$ parameters.

We have adopted a uniform prior over the set of all possible covariate subsets in both the logit and the probit case.
For each choice of the covariates, we have adopted a suitable Gaussian prior, respecting compatibility among models.

Using the importance sampling algorithm described in Section \ref{subsec:IS}, we can quickly compute, for each model, the normalizing constant in the logit case. The corresponding quantities in the probit case are available from \cite{durante2019}. Assuming that the intercept is always included, one has $2^{p-1} = 2^{54} \approx 1.8 \times 10^{16}$ possible combinations of covariates. Let $M_L$ and $M_G$ be the model subspaces in the logit and probit cases, respectively. Let $M_{L,j}$ and $M_{G,j}$ be the $j$-th logit and probit model for $j = 1, 2, \dots, 2^{54}$ respectively. Finally, let $S_L(\beta_i)$  and $S_G(\beta_i)$ be the subset of models which include the $i$-th covariate. 
For $K \in \{G, L\}$, the posterior probability of inclusion of each single covariate is
\begin{equation} \label{eq:postInclProbs}
P \left( S_K(\beta_i) \vert Y \right) = \sum\limits_{j: M_{K,j} \in S_K(\beta_i)}P(Y \vert M_{K,j})\big / {\sum\limits_{h=1}^{2^{54}} P(Y \vert M_{K,h})}\, .
\end{equation}
However, computing the marginal likelihood for all models is unfeasible in practice. Then we implemented a Gibbs variable selection technique to explore the model space and then to evaluate the expression \eqref{eq:postInclProbs}. Both for the logit and the probit models, we have run a chain of length $N=10^4$ and computed the posterior inclusion probability for each covariate as 
\begin{equation*}
P \left( S_K(\beta_i) \vert Y \right) \approx \frac{1}{N} \sum_{t =1}^N \mathbb{I}_{\{M_t \in S_K(\beta_i)\}} \, .	
\end{equation*}
Following \citet{barbieri2004}, we have selected - both in the logit and probit case - the median probability model, that is the subset of covariates such that their posterior inclusion probability is larger than $0.5$. 

We ended up with a logit model with $38$ variables and a probit model with $29$ variables; $18$ covariates are common to both median models.
An interesting by-product of our approach is that one can also make a model comparison between logit and probit links. Specifically, we have compared the median probability models under the two different links: the results show evidence in favor of the logit link, in fact, the normalizing constant of the median logit model is $3.72\times 10^{-8}$ versus a value of $3.27\times 10^{-9}$ for the median probit model, with a Bayes Factor of $11.3831$.%

%%--------------------------------------------------------------------------------------------------
\section{Conclusion}
\label{sec:conc}

Binary regression is certainly one of the most popular tools in applied statistics. 
From a Bayesian perspective, the prompt availability of an exact posterior sample for the involved parameters could be very important for producing accurate and ready-to-use summaries.

To this end, we have proposed an algorithm which, 
in the important special case of logistic regression with a Gaussian prior, can produce a posterior weighted sample of independent draws. This makes possible an easy computation of posterior summaries, including the normalizing constant, crucial to performing Bayesian model selection. 

In a more general setting, we have presented a Gibbs sampling approach, which can be used when both the link function and the prior distribution are scale mixtures of Gaussian densities. Compared with other Gibbs algorithms, we have shown, through several simulations, that the proposed approach provides a uniformly better mixing, and this gain becomes dramatic in the $p >n$ case, especially when a Dirichlet-Laplace prior is used. 
We have also stressed the fact that the algorithm does not suffer from the problem of unbalanced data, as shown in the simulation study in Section \ref{subsec:unbalanced}.

In our opinion, the gist of the paper was to introduce a self-contained algorithm for easily implementing several binary models, including the logit one, in a Bayesian framework, to provide a more informed model selection, especially when the number of covariates is large. 
Our algorithm is slower than the one described in \cite{durante2019} in the special case of a probit model with a Gaussian prior, but it allows a large class of priors and/or different link functions to be used. We also claim that an additional contribution is the possibility - at least when the prior is Gaussian - of performing
model and variable selection in a fast and automatic way.

The binary regression model can be extended in different directions and, in some cases, the extension of the method proposed here is immediate, while in other contexts it can be more problematic.
For example, in the case of the ordered multivariate logistics model, it is simple to extend our method by following the approach described in \citet{anceschi} in the probit case by 
``breaking'' the logistic CDF instead of the Normal CDF.

A more complicated issue is the obtaining of a direct extension to the multinomial logistic model because it is no longer possible to use the representation of the logistic distribution in terms of a scale mixture of Gaussian laws. 
However, it is possible to obtain a sort of extension of our approach where every marginal distribution of the response vector follows a logistic distribution, but the joint distribution, and therefore the copula structure, does not correspond to the canonical multinomial model. We have included the expression of this multivariate elliptical logistic density in the Supplementary Materials.
We are currently working in this direction to propose an efficient, sensible, and easy-to-manage dependence structure.

Another possible extension is obtained by relaxing the predictor's linearity and considering a nonparametric calibration function; our approach can be tailored to cover this case through the definition of a {\it quasi}-SUN process; we will explore these issues elsewhere.

\begin{center}
\textbf{Acknowledgements}
\end{center}
The Authors gratefully acknowledge funding from Sapienza Università di Roma, \\ grant n. RM122181612D9F93.

\section{Supplementary Materials}
The Supplementary Materials include:
\begin{description}
\item[Proofs:] all the mathematical details of the results contained in the paper. 
\item[Additional Results:] Tables and Figures related to the Simulation studies.
\item[\textbf{R} package:] a new package, \texttt{pSUN}, for implementing the methods described in this article can be downloaded at \texttt{https://github.com/paonrt/extSUN\_bbr/pSUN.git}
\end{description}

\bibliography{pSUN_ref}

\end{document}

%%%%%%%%%%%%%%%%%%%%%%%%%%%%%%%%%%%%%%%%%%%%%%%%%%%%%%%%%%%%%%%%%%%%%%%%%%%%%%
\title{\bf Supplementary Materials To \textit{An Extension of the Unified Skew-Normal Family of Distributions and Application to Bayesian Binary Regression}}
\author{Paolo Onorati \\
 Department of Statistical Sciences, University of Padua, Italy\\
 and \\
 Brunero Liseo \\
 MEMOTEF, Sapienza University of Rome, Italy}
\date{}
\maketitle

%%-----------------------------------------------------------------------------------------------
\section{Proofs of Theorems}
\label{sec:proofs}

Here we report the proofs of Theorem $1,2,3,4,5$ and $6$.

\begin{restatable}{theorem}{psun_DDFMGF}
	If $\beta\sim \psun_{p,m}\left( Q_V, \Theta, A, b, Q_W, \Omega, \xi \right)$ then the density function and the MGF of $\beta$ can be written as
	\begin{align}
		\mathrm{\emph{A}.} &\hspace{1.9mm} %\label{eq:psunDDF}
		f_\beta(\beta) = 
		\phi_{\Omega, Q_W} (\beta - \xi) \frac{
			\Phi_{\Theta, Q_V} \left( A\, \mathrm{diag}^{-\frac 12}(\Omega)
			(\beta - \xi) + b \right)
		}{
			\Psi_{Q_V, \Theta, A, Q_W, \bar{\Omega}}(b)} \, , \nonumber \\
		%%%%%%%%%%%%%%%%%%%%%%%%%%%%%%%%%%%%%%%%%%%%%%%%%%%%%%%%%%%%%%%%%%%%%%%%%%%
		\mathrm{\emph{B}.} &\hspace{1.9mm} %\label{eq:psunMGF}
		M_\beta(u) = e^{u^\prime \xi} 
		M_Z\left( \mathrm{diag}^\frac{1}{2}(\Omega)u \right) 
		\frac{
			\widetilde{\Psi}_{Q_V, \Theta, A, Q_W, \bar{\Omega}} \left(b,\mathrm{diag}^\frac{1}{2}(\Omega)u \right)
		}{
			\Psi_{Q_V, \Theta, A, Q_W, \bar{\Omega}}(b)
		} \, . \nonumber
	\end{align}
\end{restatable}

\noindent
\textbf{Proof of A:}

\noindent
It is straightforward to see that
\begin{equation*}
	f_{Z}(x) =\phi_{\bar\Omega, Q_W}(x) \, , \ f_{T}(x) =\phi_{\Theta, Q_V}(x) \, .
\end{equation*}
Let $\widetilde{\beta} = Z \vert ( T \le AZ + b )$; the density of $\widetilde{\beta}$ can be written as
\begin{align*}
	f_{\widetilde{\beta}} \left( \widetilde{\beta} \right) =& \, f_{Z} \left( \widetilde{\beta} \vert T \le AZ+b \right)\\
	=& \, f_Z \left(\widetilde{\beta} \right) \frac{\mathrm{P} \left( T \le AZ+b \vert Z = \widetilde{\beta} \right)}{\mathrm{P} \Big(T \le AZ + b \Big)}\\
	=& \, \phi_{\bar\Omega, Q_W} \left( \widetilde{\beta} \right) \frac{\Phi_{\Theta, Q_V}(A\widetilde{\beta}+b)}{\Psi_{Q_V, \Theta, A, Q_W, \bar{\Omega}}(b)} \, .
\end{align*}
However $\beta = \xi + \mathrm{diag}^{1/2}(\Omega)\widetilde{\beta}$, then
\begin{align*}
	f_{\beta}(\beta) =& \, \det\left(\mathrm{diag}^{-\frac12}(\Omega)\right) f_{\widetilde{\beta}} \left( \mathrm{diag}^{-\frac12}(\Omega)(\beta-\xi) \right) \\
	=& \, \prod_{i=1}^p \left({\Omega_{ii}^{-\frac 12}}\right) \phi_{\bar\Omega, Q_W} \left( \mathrm{diag}^{-\frac12}(\Omega)(\beta-\xi) \right) \frac{\Phi_{\Theta, Q_V}\left(A\mathrm{diag}^{-\frac12}(\Omega)(\beta-\xi)+b\right)} {\Psi_{Q_V, \Theta, A, Q_W, \bar{\Omega}}(b)} \\
	=& \, \phi_{\Omega, Q_W} (\beta - \xi) \frac{\Phi_{\Theta, Q_V}\left( A\, \mathrm{diag}^{-\frac 12}(\Omega) (\beta - \xi) + b \right)}{\Psi_{Q_V, \Theta, A, Q_W, \bar{\Omega}}(b)} \, . 
\end{align*}

\noindent
\textbf{Proof of B:}

\noindent
As before, let $\widetilde{\beta} = Z \vert T \le AZ + b$. Then
\begin{align*}
	M_{\widetilde{\beta}}(u) &= \mathrm{E} \left( e^{u^\prime \widetilde{\beta}} \right)  \\
	&= \int_{\mathbb{R}^p} e^{u^\prime x}\phi_{\bar\Omega, Q_W}(x) \frac{\Phi_{\Theta, Q_V}(Ax+b)}{\Psi_{Q_V, \Theta, A, Q_W, \bar{\Omega}}(b)} dx_1 dx_2 \cdots dx_d \\
	&= \frac{M_{Z}(u)}{\Psi_{Q_V, \Theta, A, Q_W, \bar{\Omega}}(b)} \int_{\mathbb{R}^d} \frac{e^{u^\prime x}\phi_{\bar\Omega, Q_W}(x)}{M_{Z}(u)} \Phi_{\Theta, Q_V}(Ax+b) dx_1 dx_2 \cdots dx_d \\
	&= \frac{M_{Z}(u)}{\Psi_{Q_V, \Theta, A, Q_W, \bar{\Omega}}(b)} \mathrm{E} \bigg( \mathrm{P}\left( T - A \widetilde{Z}_u \le b \vert \widetilde{Z}_u  \right) \bigg) \\
	&= M_{Z}(u) \frac{\widetilde{\Psi}_{Q_V,\Theta,A, Q_W,\bar{\Omega}}(b,u)}{\Psi_{Q_V, \Theta, A, Q_W, \bar{\Omega}}(b)} \, .
\end{align*}
Since $\beta = \xi + \mathrm{diag}^\frac{1}{2}(\Omega) \widetilde{\beta}$,
\begin{equation*}
	M_\beta(u)  = e^{u^\prime \xi}M_Z\left(\mathrm{diag}^\frac{1}{2}(\Omega)u\right) \frac{ \widetilde{\Psi}_{Q_V, \Theta, A, Q_W,      \bar{\Omega}} \left(b,\mathrm{diag}^\frac{1}{2}(\Omega)u \right)}{\Psi_{Q_V, \Theta, A, Q_W, \bar{\Omega}}(b)} \, .
\end{equation*}
%%%%%%%%%%%%%%%%%%%%%%%%%%%%%%%%%%%%%%%%%%%%%%%%%%%%%%%%%%%%%%%%%%%%%%%%%%%%%%%%
\vspace{1cm}
%%%%%%%%%%%%%%%%%%%%%%%%%%%%%%%%%%%%%%%%%%%%%%%%%%%%%%%%%%%%%%%%%%%%%%%%%%%%%%%%
\begin{restatable}{theorem}{pSUNconj}
	\label{pSUNconj}
	Consider a Bayesian LSBR model and assume that the prior for $\beta$ is 
	\begin{equation*}
		\beta \sim \psun_{p,m}( Q_{V}, \Theta, A, b, Q_W, \xi, \Omega ) \, .
	\end{equation*}
	Assume, in addition, that the link function $\Lambda(\cdot)$ has the following representation 
	\begin{equation*}
		\Lambda(x) = \int_{0}^{+\infty} \Phi \left(\frac{x}{\sqrt{v}}\right) dQ_{V^*}(v) \, .
	\end{equation*}
	Then, the posterior distribution of $\beta$ belongs to the $\psun$ family. More precisely
	\begin{equation*}
		\beta \, \vert \, (Y=y) \sim \psun_{p,m+n} 
		\left(Q_V Q_{V^*}^n,
		\begin{bmatrix}
			\Theta & 0_{m \times n} \\
			0_{n \times m} & I_{n}
		\end{bmatrix}
		\begin{bmatrix}
			A & 0_{m \times p} \\
			0_{n \times p} & B_yX\mathrm{diag}^{\frac12}(\Omega)
		\end{bmatrix}, 
		\begin{bmatrix}
			b \\
			B_y X \xi
		\end{bmatrix},
		Q_W, \xi, \Omega
		\right) \, ,
	\end{equation*}
	where we have denoted, for $[x_1 , x_2]^\prime \in \mathbb{R}^{m+n}$,
	\begin{equation*}
		Q_{V} Q_{V^*}^n \left( [x_1 , x_2]^\prime
		\right) = Q_{V}(x_1) \prod_{i=1}^n Q_{V^*} (x_{2,i}) \, .
	\end{equation*}	
\end{restatable}

\noindent
\textbf{Proof}

\noindent
The prior density of $\beta$ is:
\begin{align*}
	f_\beta(\beta) \propto \phi_{\Omega, Q_W} (y - \xi) \Phi_{\Theta, Q_V}\left( A\, \mathrm{diag}^{-\frac 12}(\Omega) (y - \xi) + b \right).
\end{align*}
The corresponding posterior density is then 
\begin{align*}
	f_\beta(\beta \vert Y=y) &\propto \\
	&\propto \Lambda_n(B_y X \beta) \phi_{\Omega, Q_W} (y - \xi) \Phi_{\Theta, Q_V}\left( A\, \mathrm{diag}^{-\frac 12}(\Omega) (y - \xi) + b \right) \\
	&= \phi_{\Omega, Q_W} (y - \xi)
	\Phi_{I_n, Q^n_{V^*}} \left( B_Y X \, \mathrm{diag}^\frac12 (\Omega) \mathrm{diag}^{-\frac12} \, (\Omega) (\beta - \xi) + B_y X \xi \right) \\
	& \hspace{0.43cm}  \Phi_{\Theta, Q_V}\left( A\, \mathrm{diag}^{-\frac 12}(\Omega) (y - \xi) + b \right) \\
	&= \phi_{\Omega, Q_W} (y - \xi) 
	\Phi_{{\Theta}^\ast,
		Q_{V_0}Q^n_{V^*}
	}
	\left(
		\begin{bmatrix}
			A \\
			B_y X \; \mathrm{diag}^\frac12 (\Omega)
		\end{bmatrix}
		\mathrm{diag}^{-\frac12} (\Omega) ( \beta - \xi) + 
		\begin{bmatrix}
			b \\
			B_y X \xi
		\end{bmatrix}
	\right) \, ,
\end{align*}
where ${\Theta}^\ast= \begin{bmatrix}
	\Theta & 0_{m,n} \\
	0_{n,m} & I_n
\end{bmatrix}$.
%%%%%%%%%%%%%%%%%%%%%%%%%%%%%%%%%%%%%%%%%%%%%%%%%%%%%%%%%%%%%%%%%%%%%%%%%%%%%%%%
\vspace{1cm}
%%%%%%%%%%%%%%%%%%%%%%%%%%%%%%%%%%%%%%%%%%%%%%%%%%%%%%%%%%%%%%%%%%%%%%%%%%%%%%%%
\begin{restatable}{theorem}{postLogisKolmo}
	\label{postLogisKolmo}
	Let $V_i \sim \LK(\cdot)$ and $T_i \vert V_i \sim N(0, V_i)$.
	Then:
	\begin{align*}
		\mathrm{\emph{A}.} &\hspace{1.9mm} M_{V_i}(u \vert T_i = t) = \mathrm{E}(e^{u V_i} \vert T_i = t) = e^{\abst} ( 1 + e^{-\abst} ) \sum_{k = 1}^{+\infty} (-1)^{k+1} k^2 \frac{ \exp \left( -\abst \sqrt{k^2 - 2u} \right) }{ \sqrt{k^2 - 2u} } \, , \\
		\mathrm{\emph{B}.} &\hspace{1.9mm} \mathrm{E}(V_i \vert T_i = t) = ( 1 + e^{-\abst} ) \left( \abst + (1 + e^{\abst}) \log(1 + e^{-\abst}) \right) \, .
	\end{align*}
\end{restatable}

\noindent
\textbf{Proof of A:}

\noindent
Notice that $[ V_i \vert T_i = t ] \overset{d}{=} [ V_i \vert T_i = -t ]$, therefore without loss of generality we can assume $t > 0$. Thus
\begin{align}
	M_{V_i}(u \vert 	T_i = t) &= \int_{0}^{+\infty} e^{uv} f_{V_i}(v \vert T_i = t) dv  \nonumber\\
	&= \frac{ e^t (1+e^{-t})^2 }{ \sqrt{2 \pi} } \int_{0}^{+\infty} \sum_{k=1}^{\infty} (-1)^{k+1} k^2 v^{-\frac12} \exp \left( -\frac12 \Big(  (k^2-2u)v + \frac{t^2}{v} \Big) \right)dv \, . \label{eq:b.1}
\end{align}
Let $a_k(v) = k^2 v^{-\frac12} \exp \left( -\frac12 \Big(  (k^2-2u)v + \frac{t^2}{v} \Big) \right)$, then it is straightforward
\begin{align*}
	a_k(v) &> 0 \ \forall \, v > 0 \, , \\
	\exists \, k^\ast \in \mathbb{N} : a_k(v) &> a_{k+1}(v) \ \forall \, k > k^\ast \, \mathrm{and} \ \forall \, v > 0 \, .
\end{align*}
The above properties imply that we can use the Leibniz criterion in order to verify the uniform convergence of $ \sum_{k = 1}^{+\infty}(-1)^{k+1} a_k(v)$, thus
\begin{align*}
	\frac{\d \log a_k(v)}{\d v} &= -\frac{1}{2 v} -\frac12 (k^2 - 2u) + \frac{t^2}{2 v^2} \, , \\
	\arg \sup_{v > 0} a_k(v) &= \frac{ \sqrt{1+4(k^2-2u)}-1 }{2 (k^2 - 2u)}
\end{align*}
and
\begin{equation*}
	\sup_{v > 0} a_k(v) = \frac{ k^2 \sqrt{2 k^2 - 2u} }{ \sqrt{ \sqrt{1 + 4 (k^2-2u)} - 1 } } \exp \left( -\frac12 \bigg( k^2 \frac{  \sqrt{1 + 4 (k^2-2u)} - 1  }{ 2 (k^2 - 2u) } + t^2 \frac{ 2(k^2-2u) }{ \sqrt{1+4(k^2-2u)} - 1 } \bigg) \right) \, ,
\end{equation*}
assume $\vert u \vert < 1/4$, then
\begin{align*}
	\sup_{v > 0} a_k(v) &\le 2 k^3 \exp \left( \frac14 - \frac{k}{ 2\sqrt{2} } \right) \, , \\
	\lim_{k \to +\infty} \sup_{v > 0} a_k(v) &\le 2 \lim_{k \to +\infty} k^3 \exp \left( \frac14 - \frac{k}{ 2\sqrt{2} } \right) = 0 \, ,
\end{align*}
so $\sum_{k = 1}^{+\infty}(-1)^{k+1} a_k(v)$ converges uniformly and this implies that we can change the order between integral and sum in \eqref{eq:b.1}.
Therefore
\begin{align*}
	M_{V_i}(u \vert T_i = t) &= \frac{e^t (1+e^t)^2}{ \sqrt{2 \pi} } \sum_{k = 1}^{+\infty} (-1)^{k+1} k^2 \int_{0}^{+\infty} v^{-\frac12} \exp \left( -\frac12 \Big( (k^2-2u)v + \frac{t^2}{v} \Big) \right) dv \\
	&= \frac{e^t (1+e^t)^2}{ \sqrt{2 \pi} } \sum_{k = 1}^{+\infty} (-1)^{k+1} k^2 2 \, \mathrm{K}_\frac12 \left( t \sqrt{k^2-2u} \right) \left( \frac{k^2-2u}{t^2} \right)^{-\frac14} \, ,
\end{align*}
where $\mathrm{K}_s(\cdot)$ is a modified Bessel function of second kind with order $s$; it is well know that
\begin{equation*}
	\mathrm{K}_\frac12(x) = \sqrt{ \frac{\pi}{2x} } e^{-x} \, ,
\end{equation*}
therefore we obtain
\begin{align*}
	M_{V_i}(u \vert T_i = t) &= \frac{e^t (1+e^t)^2}{ \sqrt{2 \pi} } \sum_{k = 1}^{+\infty} (-1)^{k+1} 2 k^2 t^\frac12 (k^2-2u)^{-\frac14} \sqrt{ \frac{\pi}{2} } \left( t \sqrt{k^2-2u} \right)^{-\frac12} \exp \left( -t \sqrt{k^2-2u} \right) \\
	&= e^t (1+e^{-t}) \sum_{k = 1}^{+\infty} (-1)^{k+1} k^2 \frac{ \exp \left( -t \sqrt{k^2-2u} \right) }{ \sqrt{k^2 -2u} } \, .
\end{align*}

\noindent
\textbf{Proof of B:}

\noindent
As before, without loss of generality we can assume $t > 0$. So
\begin{equation} \label{eq:b.2}
	\frac{\d M_{V_i}(u \vert T = t)}{\d u} = e^t (1 + e^{-t})^2 \frac{\d}{\d u} \left[ \sum_{k = 1}^{+\infty} k^2 \frac{ \exp \left( -t \sqrt{k^2-2u} \right) }{ \sqrt{k^2 -2u} } \right] \, ,
\end{equation}
let 
\begin{align*}
	a_k(u) &= k^2 \frac{ \exp \left( -t \sqrt{k^2-2u} \right) }{ \sqrt{k^2 -2u} } \, , \\
	b_k(u) &= \frac{\d a_k(u)}{\d u} = \frac{k^2}{k^2-2u} \left( t + \frac{1}{ \sqrt{k^2-2u} } \right) \exp \left( -t \sqrt{k^2-2u} \right) \, ,
\end{align*}
as before assume $\vert u \vert < 1/4$, after some algebra one can show that
\begin{align*}
	b_k(u) &> 0 \ \forall \, \vert u \vert < \frac14 \, , \\
	\exists \, k^\ast \in \mathbb{N} : b_k(u) &> b_{k+1}(u) \ \forall \, k > k^\ast \, \mathrm{and} \ \forall \, \vert u \vert < \frac14 \, ,
\end{align*}
i.e. we can still use the Leibniz criterion in order to asses the uniform convergence of $\, \sum_{k = 1}^{+\infty} (-1)^{k+1} b_k(u)$. Therefore
\begin{align*}
	\sup_{\vert u \vert < 1/4} b_k(u) &= \sup_{\vert u \vert < 1/4} \frac{k^2}{k^2-2u} \left( t + \frac{1}{ \sqrt{k^2-2u} } \right) \exp \left( -t \sqrt{k^2-2u} \right) \\
	& \le \frac{k^2}{k^2-\frac12} \left( t + \frac{1}{ \sqrt{k^2-\frac12} } \right) \exp \left( -t \sqrt{k^2-\frac12} \right) \, ,
\end{align*}
so
\begin{equation*}
	\lim_{k \to +\infty} \sup_{\vert u \vert < 1/4} b_k(u) \le \lim_{k \to +\infty} \frac{k^2}{k^2-\frac12} \left( t + \frac{1}{ \sqrt{k^2-\frac12} } \right) \exp \left( -t \sqrt{k^2-\frac12} \right) = 0
\end{equation*}
so $\sum_{k = 1}^{+\infty}(-1)^{k+1} b_k(u)$ converges uniformly and this implies that we can change the order between integral and sum in \eqref{eq:b.2}.
Therefore
\begin{align*}
	\frac{\d M_{V_i}(u \vert T = t)}{\d u} &= e^t (1+e^{-t})^2 \sum_{k=1}^{+\infty} (-1)^{k+1} \frac{k^2}{k^2-2u} \left( t + \frac{1}{ \sqrt{k^2-2u} } \right) \exp \left( -t \sqrt{k^2-2u} \right) \, , \\
	\left. \frac{\d M_{V_i}(u \vert T = t)}{\d u} \right \vert_{u = 0} &= \mathrm{E}(V_i \vert T = t) = e^t (1+e^{-t})^2 \sum_{k=1}^{+\infty} (-1)^{k+1} \left( t + \frac1k \right) e^{-tk} \, ,
\end{align*}
and we obtain
\begin{equation} \label{eq:b.3}
	\mathrm{E}(V_i \vert T = t) = e^t (1+e^{-t})^2 \left( t \sum_{k=1}^{+\infty}(-1)^{k+1} e^{-tk} + \sum_{k=1}^{+\infty}(-1)^{k+1} \frac{ e^{-tk} }{k} \right) \, .
\end{equation}
Consider separately the first sum of \eqref{eq:b.3},
\begin{align*}
	\sum_{k=1}^{+\infty}(-1)^{k+1} e^{-tk} &= \sum_{h=0}^{+\infty}(-1)^{h} e^{-t(h+1)} \\
	&= e^{-t} \sum_{h=0}^{+\infty} \left(-1 e^{-t} \right)^h = \frac{ e^{-t} }{ 1+e^{-t} } \, ,
\end{align*}
Consider separately the second sum of \eqref{eq:b.3}, i.e
\begin{equation*}
	c(t) = \sum_{k=1}^{+\infty}(-1)^{k+1} \frac{ e^{-tk} }{k}
\end{equation*}
and it is well know that $c(0) = \log 2$, furthermore it is straightforward to see that the series of $c(t)$ converges uniformly; then
\begin{align*}
	\frac{\d c(t)}{\d t} &= \sum_{k=1}^{+\infty}(-1)^{k} e^{-tk} \\
	&= -\sum_{k=1}^{+\infty}(-1)^{k+1} e^{-tk} \, ,
\end{align*}
and this is equal to the first sum of \eqref{eq:b.3} up to a sign, so
\begin{align*}
	\frac{\d c(t)}{\d t} = - \frac{ e^{-t} }{ 1+e^{-t} } \, ;
\end{align*}
then we obtain
\begin{align*}
	c(t) &= c(0) + \int_{0}^{t} \d c(s) \\
	&= \log 2 - \int_{0}^{t} \frac{ e^{-s} }{ 1+e^{-s} } \d s = \log(1+e^{-t}) \, .
\end{align*}
Finally
\begin{align*}
	\mathrm{E}(V_i \vert T_i = t) &= e^t (1+e^{-t})^2 \left( \frac{ t e^{-t} }{ 1+e^{-t} } + \log(1+e^{-t}) \right) \\
	&= (1+e^{-t}) \left( t + (1 + e^{t}) \log(1 + e^{-t}) \right) \, .
\end{align*}
%%%%%%%%%%%%%%%%%%%%%%%%%%%%%%%%%%%%%%%%%%%%%%%%%%%%%%%%%%%%%%%%%%%%%%%%%%%%%%%%
\vspace{1cm}
%%%%%%%%%%%%%%%%%%%%%%%%%%%%%%%%%%%%%%%%%%%%%%%%%%%%%%%%%%%%%%%%%%%%%%%%%%%%%%%%
\begin{restatable}{theorem}{accRej_plk}
	\label{accRej_plk}
	Let $V_i \sim \LK(\cdot)$, $T_i \vert V_i \sim N(0, V_i)$, $\widetilde{V} \sim \igamma(\alpha, \pi^2 / 2)$ and $\widetilde{T} \vert \widetilde{V} \sim N(0, \widetilde{V})$; set $\alpha > 3 / 2$.
	Then the ratio $f_{V_i}(v \vert T_i = t) / f_{\widetilde{V}}(v \vert \widetilde{T} = t)$ is bounded above by
	\begin{equation}
		\label{accRej_bound}
		M^\ast \frac{ f_{\widetilde{T}}(t) }{ f_{T_i}(t) } \, ,
	\end{equation}
	with $M^\ast = \max \Big( \max\limits_{0 < v \le v^\ast} \delta_1(v) \, , \, \max\limits_{v > v^\ast} \delta_2(v) \Big) $, $v^\star \in (1/2, 18 \pi^2 / 11)$,
	\begin{align*}
		\begin{bmatrix}
			\delta_1(v) \\
			\delta_2(v) 
		\end{bmatrix} &= \left\{
		\begin{alignedat}{2}
			& \frac{\sqrt{2 \pi^5} \Gamma(\alpha)} {(\pi^2/2)^\alpha} v^{\alpha - \frac 32} \quad &\mathrm{if}& \quad 0 < v \le v^\ast \\
			& \frac{\Gamma(\alpha)} {(\pi^2/2)^\alpha} v^{\alpha+1} \exp \left( \frac {\pi^2}{2 v} - \frac v 2 \right) \quad &\mathrm{if}& \quad v > v^\ast
		\end{alignedat}
		\right. \, , \\
		\arg \max_{0 < v \le v^\ast} \delta_1(v) &= v^\ast \, , \\
		\arg \max_{v > v^\ast} \delta_2(v) &= \left\{
	\begin{alignedat}{2}
		&1 + \alpha + \sqrt{(1+\alpha)^2 - \pi^2} \quad &\mathrm{if} \quad \alpha \ge \pi -1 \\
		&v^\ast &\mathrm{otherwise} \qquad \quad \ \,
	\end{alignedat}
		\right. \, .
	\end{align*}
\end{restatable}

\noindent
\textbf{Proof:}

\noindent
Notice that 
\begin{equation*}
	f_{V_i}(v \vert T_i = t) = \frac{f_{T_i}(t \vert V_i = v) f_{V_i}(v) }{f_{T_i}(t)} = \frac{ \varphi(t /\sqrt{v}) \lk(v) }{\sqrt{v} f_{T_i}(t)} \, ,
\end{equation*}
where $f_{T_i}(\cdot)$ is the density of the standard logistic distribution. As a proposal density, we consider $\widetilde{V} \vert \widetilde{T}$, that is
\begin{equation*}
	f_{\widetilde V}(v \vert \widetilde T = t) = \frac{f_{ \widetilde T}(t \vert \widetilde V = v) f_{\widetilde V}(v) }{f_{\widetilde T}(t)} = \frac{\varphi(t /\sqrt{v}) f_{\widetilde V}(v) }{\sqrt{v} f_{\widetilde T}(t)} \, ,
\end{equation*}
with $\widetilde{V} \sim \igamma(\alpha, \gamma)$ so $\widetilde{V} \vert \widetilde{T} = t \sim \igamma(\alpha + 1/2, \gamma + t^2/2 )$, and
\begin{equation*}
	f_{\widetilde T}(t) = \frac{\Gamma(\frac{2 \alpha + 1}{2})}{\Gamma(\alpha)\sqrt{2 \pi \gamma}} \left( 1 + \frac{t^2}{2 \gamma} \right)^{-\frac{2 \alpha + 1}{2}} \, .
\end{equation*} 
Then
\begin{equation*}
	\frac{f_{V_i}(v \vert T_i = t)}{f_{\widetilde V}(v \vert \widetilde T = t)} = \frac{\lk(v)}{f_{\widetilde V}(v)} \frac{f_{\widetilde T}(t)}{f_{T_i}(t)} \, .
\end{equation*}
As a consequence, only the first ratio depends on $v$, and we need to find an upper bound $M^\ast$ such that
\begin{equation*}
	\frac{\lk(v)}{f_{\widetilde V}(v)} \leq M^\ast \, , \forall v > 0 \, .
\end{equation*}
For any $v^\ast \in \mathbb{R}$, one can see that 
\begin{equation*} 
	%\label{eq:ratioAR}
	\frac{\lk(v)}{f_{\widetilde{V}}(v)} = \left\{
	\begin{alignedat}{2}
		& \frac{\sqrt{2 \pi} \Gamma(\alpha)} {\gamma^{\alpha}} v^{\alpha - \frac 32} \sum_{j=1}^{+\infty} \left( (2j-1)^2 \pi^2 - v \right) \exp \left( \frac \gamma v - \frac{(2j-1)^2 \pi^2}{2 v} \right) \quad &\mathrm{if}& \quad 0 < v \le v^\ast \\
		& \frac{\Gamma(\alpha)} {\gamma^\alpha} v^{\alpha+1} \sum_{j=1}^{+\infty} (-1)^{j-1} j^2 \exp\left( \frac \gamma v - \frac{j^2 v}{2} \right) \quad &\mathrm{if}& \quad v > v^\ast
	\end{alignedat}
	\right. \, .
\end{equation*}
Thus, we set $\gamma = \pi^2/2$ and, with simple algebra, it is possible to show that the above expression is bounded above by
\begin{equation*} 
	%\label{eq:boundAR}
	\begin{bmatrix}
		\delta_1(v) \\
		\delta_2(v) 
	\end{bmatrix} = \left\{
	\begin{alignedat}{2}
		& \frac{\sqrt{2 \pi^5} \Gamma(\alpha)} {(\pi^2/2)^\alpha} v^{\alpha - \frac 32} \quad &\mathrm{if}& \quad 0 < v \le v^\ast \le \frac{18 \pi^2}{11}\\
		& \frac{\Gamma(\alpha)} {(\pi^2/2)^\alpha} v^{\alpha+1} \exp \left( \frac {\pi^2}{2 v} - \frac v 2 \right) \quad &\mathrm{if}& \quad v > v^\ast \ge \frac{1}{2}
	\end{alignedat}
	\right. \, .
\end{equation*} 
Thus $	M^\ast = \max \Big( \max_{0 < v \le v^\ast} \delta_1(v) \, , \, \max_{v > v^\ast}\delta_2(v) \Big)$. We restrict $\alpha$ to be larger than $3/2$, so it is straightforward
\begin{equation*}
	\arg \max_{0 < v \le v^\ast} \delta_1(v) = v^\ast \, . \\
\end{equation*}
Furthermore, it is easy to show
\begin{equation*}
	\arg \max_{v > v^\ast} \delta_2(v) = \left\{
	\begin{alignedat}{2}
		&1 + \alpha + \sqrt{(1+\alpha)^2 - \pi^2} \quad &\mathrm{if} \quad \alpha \ge \pi -1 \\
		&v^\ast &\mathrm{otherwise} \qquad \quad \ \,
	\end{alignedat}
	\right. \, .
\end{equation*}
%%%%%%%%%%%%%%%%%%%%%%%%%%%%%%%%%%%%%%%%%%%%%%%%%%%%%%%%%%%%%%%%%%%%%%%%%%%%%%%%
\vspace{1cm}
%%%%%%%%%%%%%%%%%%%%%%%%%%%%%%%%%%%%%%%%%%%%%%%%%%%%%%%%%%%%%%%%%%%%%%%%%%%%%%%%
\begin{restatable}{theorem}{sutPDF}
	\label{sut}
	If $\zeta_1 \sim \sut_{p,n}(\nu, \Theta, A, b, \xi, \Omega)$ then its density function is
	\begin{equation*}
		f_{\zeta_1}(x) = t_{\nu, \Omega}(x-\xi) \frac{
			T_{\nu+p, -b, \Theta} \left( \sqrt{ \frac{\nu+p}{\nu + (x-\xi)^\prime \, \Omega^{-1} \, (x-\xi)} } A \, \mathrm{diag}^{ -\frac{1}{2} }(\Omega) (x-\xi) \right)
		} {\Phi_{\Theta + A \, \OmegaBar A^\prime}(b)} \, .
	\end{equation*}
	Where $t_{\nu, \Omega}(\cdot)$ is the density function of a Student-$t$ random variable with $\nu$ degrees of freedom and scale matrix $\Omega$ and $T_{\nu, b, \Theta}(\cdot)$ is the CDF of a random variable $R_1$ with the following stochastic representation:
	\begin{align} 
		\label{eq:mnct}
		\begin{split}
			&R_1 = \sqrt{S}(R_0 + b) \, , \\
			&S \sim \igamma \left( \frac{\nu}{2}, \frac{\nu}{2} \right) \indep R_0 \sim N_n(0, \Theta) \, .
		\end{split}
	\end{align}
\end{restatable}

\noindent
\textbf{Proof:}

\noindent
Using the definition of a $\sut$ random variable, it is straightforward to obtain the following stochastic representation:
\begin{equation*}
	\zeta_1 = \xi + \sqrt{S} \, \mathrm{diag}^\frac{1}{2}(\Omega) Z \vert T \le A \, Z + b \, ;
\end{equation*}
notice that $(T \le A \, Z + b)$ is equivalent to $(\sqrt{S}T \le \sqrt{S} A \, Z + \sqrt{S}b)$, so we define
\begin{equation*}
	\widehat{T} = \sqrt{S} T \, , \ \widehat{Z} = \sqrt{S} Z \, , \ \widehat{b} = \sqrt{S} b
\end{equation*}
and we obtain another stochastic representation, that is
\begin{equation}
	\label{newDefSUT}
	\zeta_1 = \xi + \mathrm{diag}^\frac{1}{2}(\Omega) \widehat{Z} \vert \widehat{T} \le A \, \widehat{Z} + \widehat{b} \, .
\end{equation} 
Then, we compute the density of $\zeta_1$ using expression \eqref{newDefSUT}; indeed we have
\begin{align*}
	f_{\zeta_1}(x) &= \prod_{i=1}^{p}(\Omega_{i,i})^{-\frac{1}{2}} f_{\widehat{Z}} \left( \diagInvOHf(\Omega) (x - \xi) \right) \frac{ \mathrm{P} \left( \widehat{T} \le A \, \widehat{Z} + \widehat{b} \vert \widehat{Z} = \mathrm{diag}^{-\frac{1}{2}}(\Omega)(x-\xi) \right) } { \mathrm{P} \left( \widehat{T} \le A \widehat{Z} + \widehat{b} \right) } \\
	&= t_{\nu, \Omega}(x-\xi) \frac{ \mathrm{P} \left( \widehat{T} \le A \, \widehat{Z} + \widehat{b} \vert \widehat{Z} = \mathrm{diag}^{-\frac{1}{2}}(\Omega)(x-\xi) \right)} {\Phi_{\Theta + A \, \OmegaBar \, A^\prime}(b)} \, ,
\end{align*}
where $t_{\nu, \Omega}(\cdot)$ is the density function of a Student-$t$ random variable with $\nu$ degrees of freedom and scale matrix $\Omega$. It is worth noticing that the normalizing constant of the above density is the same as the one of the associated $\sun$, say $\zeta_0$. Then we only need to compute the probability at the numerator. To this end, note that 
\begin{align*}
	S \, \vert \, \widehat{Z} = \widehat{z} &\sim \igamma \left( \frac{\nu+p}{2}, \frac{\nu + \widehat{z}^\prime \, \OmegaBar^{-1} \, \widehat{z}}{2} \right) \, , \\
	\frac{ \nu+p }{ \nu + \widehat{z}^\prime \, \OmegaBar^{-1} \, \widehat{z} } \, S \, \vert \, \widehat{Z} = \widehat{z} &\sim \igamma \left( \frac{\nu+p}{2}, \frac{\nu+p}{2} \right) .
\end{align*} 
Then
\begin{align*}
	&\mathrm{P} \left( \widehat{T} \le A \, \widehat{Z} + \widehat{b} \, \vert \, \widehat{Z} = \mathrm{diag}^{-\frac{1}{2}}(\Omega)(x-\xi) \right) \\
	&= T_{\nu+p, -b, \Theta} \left( \sqrt{ \frac{\nu+p}{\nu + (x-\xi)^\prime \, \Omega^{-1} \,(x-\xi)} } A \, 	\mathrm{diag}^{ -\frac{1}{2} }(\Omega) (x-\xi) \right) \, ,
\end{align*}
where $T_{\nu_0, b_0, \Theta_0}(\cdot)$ is the CDF of a random variable $R_1$ with the following stochastic representation:
\begin{align} 
	\label{eq:mnct2}
	\begin{split}
		&R_1 = \sqrt{S}(R_0 + b_0) \, , \\
		&S \sim \igamma \left( \frac{\nu_0}{2}, \frac{\nu_0}{2} \right) \indep R_0 \sim N_n(0, \Theta_0) \, .
	\end{split}
\end{align}
%%%%%%%%%%%%%%%%%%%%%%%%%%%%%%%%%%%%%%%%%%%%%%%%%%%%%%%%%%%%%%%%%%%%%%%%%%%%%%%%
\vspace{1cm}
%%%%%%%%%%%%%%%%%%%%%%%%%%%%%%%%%%%%%%%%%%%%%%%%%%%%%%%%%%%%%%%%%%%%%%%%%%%%%%%%
\begin{restatable}{theorem}{ISgood}
	\label{ISgood}
	In an LSBR model. If the posterior distribution is a $\psun_{p,n}(Q_V, I_n, A, b, \\ \mathbb{I}_{ \{ w \ge 1_p \} }, \xi, \Omega)$ and it is used, as importance density, a $\sut_{p, n}(\nu_n, I_n, \diagInvOHf(\Vfix)A, \diagInvOHf \\ (\Vfix)b, \xi^\star, \Omega)$, then
the importance weights are bounded above for every choice of $\xi^\star$, $\nu_n < +\infty$, $\Vfix$, and for every sample size $n$.
\end{restatable}

\noindent
\textbf{Proof:} 

\noindent
Both the target and the importance density are strictly positive and bounded on their support $\mathbb{R}^p$. 
Then the ratio could diverge only if the tails of the importance density are thinner than the target.

However, if $\beta \sim \sut_{p, n}(\nu, \Theta, A, b, \xi, \Omega)$, it can be shown that there exists an $m$-dimensional vector $\widehat{U} = \widehat{T} - A \widehat{Z}$ such that
\begin{equation}
	\label{sutGivenU}
	\beta \vert \widehat{U} \sim \textrm{Stud} \left( \nu + m, \; \xi + \diagOHf(\Omega) \OmegaBar A^\prime (\Theta + A \OmegaBar A^\prime)^{-1} \widehat{U}, \; \Omega(\widehat{U}) \right) \,
\end{equation}
where $\textrm{Stu}(\nu_0, \xi_0, \Omega_0)$ is a Student-$t$ random variable with $\nu_0$ degrees of freedom, location $\xi_0$ and scale $\Omega_0$; also
\begin{equation*}
	\Omega(\widehat{U}) = \frac{\nu + \widehat{U}^\prime (\Theta + A \OmegaBar A^\prime)^{-1} \widehat{U}  }{\nu + m} \Big( \OmegaBar - \OmegaBar A^\prime (\Theta + A \OmegaBar A^\prime)^{-1} A \OmegaBar \Big) \, .
\end{equation*}
Then, if $k \ge (\nu + m)$, 
$\mathrm{E}(\beta^k \vert U)$ is not defined and, a fortiori,   $\mathrm{E}(\beta^k)$ is also undefined. Therefore, the $\sut$ density function multiplied by $\beta^k$  is not integrable as $||\beta|| \rightarrow +\infty$ for $k \ge (\nu + m)$. 
On the other hand, we know from Theorem 1, that the target density has all finite moments, so the target multiplied by $\beta^k$ times is integrable for all positive $k$. Therefore, there exists a constant $h$ such that, if $||\beta|| > h$, the target density is smaller than the importance density; this guarantees that the importance weight do not diverge as $||\beta|| \rightarrow +\infty$.
\section{Other Results}

\subsection{Gibbs sampler step for \texorpdfstring{$\boldsymbol{Z, T \vert W, V}$}{Z, T | W, V}}

We describe a Gibbs sampler step for a $\psun$ random variable. In particular, we focus on updating $(Z, T)$ conditionally on $(W, V)$. Recall that if $\beta \sim \psun_{p,m} \left( Q_V, \Theta, A, b, Q_W, \xi, \Omega \right)$, then its stochastic representation is
\begin{equation} \label{eq:sto_pSUN}
	\beta = \xi +  \diagOHf (\Omega) Z \vert T \leq A Z + b \, ,
\end{equation}
where $Z$ and $T$ are independent and, conditionally on $(W, V)$,
\begin{equation*}
	Z  \sim N_p(0, \OmegaBar_W ) \indep 
	T \sim N_m(0, \Theta_V)\,;\quad \, 
	W \sim Q_W(\cdot) \indep  V \sim Q_V(\cdot).
\end{equation*}
Thus, setting $\varepsilon = T - AZ$, expression \eqref{eq:sto_pSUN} becomes
\begin{equation} \label{eq:sto_pSUN_eps}
	\beta = \xi +  \diagOHf (\Omega) Z \vert \varepsilon \leq b \, .
\end{equation}
However, conditionally on $(W,V)$, $\varepsilon \sim N_p(0, \Theta_V + A \OmegaBar_W A^\prime)$,  and $\cov(Z, \varepsilon \vert W, V) = - \OmegaBar_W A^\prime$. Therefore, we obtain
\begin{equation} \label{eq:cond_Z}
	Z \vert \varepsilon, W, V \sim N_p \Big( -\OmegaBar_W A^\prime \big( \Theta_V + A \OmegaBar_W A^\prime \big)^{-1} \varepsilon \, , \: \OmegaBar_W - \OmegaBar_W A^\prime \big( \Theta_V + A \OmegaBar_W A^\prime \big)^{-1} A \OmegaBar_W \Big) \, .
\end{equation}
Thus, setting $H_\mu = -\OmegaBar_W A^\prime \big( \Theta_V + A \OmegaBar_W A^\prime \big)^{-1}$, $H_\Sigma = (I + H_\mu A^\prime) \OmegaBar_W$ and using expression \eqref{eq:sto_pSUN_eps}, we sample $\varepsilon \sim \mathrm{TN}_m(-\infty, b, 0, H_\Sigma)$. Finally, we sample $Z$ using expression \eqref{eq:cond_Z}. Furthermore, $\varepsilon = T - A Z$ and, after obtaining $\varepsilon$ and $Z$, the value of $T$ is automatically given by $T = \varepsilon + A Z$.

\subsection{The density of a Kolmogorov random variable}

Here we report the expression of the density 
of the Kolmogorov random variable used in \S~4.

$$
\qkol(v)= \begin{cases}
 8x \sum\limits_{k=1}^{+\infty} (-1)^{k-1} k^2 \exp \left(-2k^2 v^2 \right) & v \ge v_0 \\
\frac{\sqrt{2\pi}}{v^2} \sum\limits_{k=1}^{+\infty} \left( \frac{(2k-1)^2 \pi^2}{4v^2} -1 \right) \exp \left( -\frac{(2k-1)^2 \pi^2}{8v^2} \right)  & 0 < v \le v_0 \,
 \end{cases},
$$
where $v_0>0$ is an arbitrary finite value. 

\subsection{Logit model as a mixture of Probit models}
A general version of a probit model can be written in the following way:
\begin{align*}
Y_i &= \mathbb{I}_{ \{ U_i \le 0 \} } \, , \\
U_i &\overset{ind}{\sim} N(-X^\prime_i \beta, V_i) \, , \, i = 1,2,\dots, n \, ,
\end{align*}
where $\mathbb{I}_{\{A\}}$ is the indicator function of the set $A$, $X_i^\prime$ is the $i$-th row of $X$ and $\beta, U_1, U_2, \dots, U_n,$ $ V_1, V_2, \dots, V_n$ are unknown and unobserved; thus the model is not identifiable. In the most common version of the probit model the lack of identifiability is circumvented by setting $V_1 = V_2 = \dots = V_n = 1$; however, from a practical perspective, in the logit model the non-identifiability is solved by putting a distribution, say $V_i \overset{iid}{\sim} \LK(\cdot)$. Then, a logit model can be viewed as a mixture of a general form of probit models, a single component of this mixture is characterized by the vector $V = \{V_1, V_2, \dots, V_n\}$, which we denote with $M(V)$. For the sake of clarity assume a Gaussian prior; then our proposed Gibbs can be simplified in the following way:
\begin{align*}
	\mathrm{sample}& \hspace{0.25cm} M_t(V) \, \vert \, \beta_{t-1}, U_{t-1} \, , \\
	\mathrm{sample}& \hspace{0.25cm} \beta_{t}, U_{t} \, \vert \, M_t(V) \, .
\end{align*}
Now consider $2$ subsequent steps, say $\beta_{t}, U_{t} \, \vert \, M_t(V)$ and $\beta_{t+1}, U_{t+1} \, \vert \, M_{t+1}(V)$. In particular, the marginal distributions of $\beta_{t} \, \vert \, M_t(V)$ and $\beta_{t+1} \, \vert \, M_{t+1}(V)$ represent two independent draws from two different $\sun$ distributions; in other terms, they can be viewed as simulations from the posterior distributions of different probit models. On the other hand, within a Polya-Gamma approach, the conditional distribution of $\beta$ is always Gaussian. 
In principle, the posterior distribution arising from 
a probit model, that is a $\sun$ distribution, should be more similar to the posterior distribution arising from a logistic regression model, i.e. a $\psun$ distribution, compared to a Gaussian one, indeed, the $\sun$ distribution retains a similar selection mechanism of the $\psun$ distribution that maintains the skewness.
This implies that, in the case of logistic regression, 
the $\psun$ posterior is typically less correlated with the logistic Kolmogorov random variables compared to what occurs in the Polya-Gamma algorithm. 
This also motivates the use of a $\sut$ density instead of a symmetric Student-$t$ in the importance sampling algorithm. 

\subsection{The multivariate elliptical logistic density}
A multivariate elliptical logistic density can be obtained as a scale mixture of Gaussian density. Let $T \vert V \sim N_p(0, V \Theta)$ and $V\sim \LK(\cdot)$. Then the marginal density of $T$ has two equivalent expressions.
The former is 
\begin{equation*}
	f_T(t) = \frac{2 (t^\prime \Theta^{-1} t)^{(2-p)/4}}{\sqrt{(2\pi)^p \det(\Theta)}} \sum_{h=1}^{\infty} (-1)^{h+1} h^{p/2 +1} K_{p/2-1}\left(h \sqrt{t^\prime \Theta^{-1} t}\right) \, .	
\end{equation*}
The latter is
\begin{equation*}
f_T(t) = \frac{2 \Gamma \big( \frac{p+1}{2} \big)}{\sqrt{\pi^{p-1} \det(\Theta)}} \sum_{h=1}^{\infty} \frac{(2h-1)^2 \pi^2 p - t^\prime \Theta^{-1} t}{ \big( t^\prime \Theta^{-1} t + (2h-1)^2 \pi^2 \big)^{\frac{p+3}{2}}} \, .
\end{equation*}

\section{Examples: Additional Results }
\label{sec:addRes}

Here we report some additional results and graphics related to simulation studies and empirical applications.

\subsection{Simulation Study: Coverage Analysis}
Tables \ref{IScoverage_tab_intercept} and \ref{IScoverage_tab_notIntercept} report the frequentist coverage for the intercept and the other parameters respectively, using the $\psun$-IS algorithm in the logit case. For the same simulation experiment, Tables \ref{IScoverage_ESS}, and \ref{IScoverage_time} report summaries of the \textit{effective sample size}, and the computational time respectively.

\begin{table}[ht]
	\caption{ Coverage Analysis of $\psun$-IS: Frequentist Coverage vs Theoretical Coverage for the Intercept, in the logistic regression. Different Combinations of Sample Size $n$ and Number of Parameters $p$. Number of simulations $N = 10^4$. }
	\label{IScoverage_tab_intercept}
	\begin{center}
		\begin{tabular}{ | c | c c | c c | c c | }
			\hline
			\multirow{2}{*}{Theoretical Coverage} & \multicolumn{2}{c|}{$n = 50$} & \multicolumn{2}{c|}{$n = 100$} & \multicolumn{2}{c|}{$n = 200$} \\
			& $p = 500$ & $p = 1000$ & $p = 500$ & $p = 1000$ & $p = 500$ & $p = 1000$ \\ 
			\hline
		0.05 & 0.0480 & 0.0518 & 0.0522 & 0.0484 & 0.0504 & 0.0483 \\ 
		0.10 & 0.0998 & 0.0995 & 0.1045 & 0.1047 & 0.1053 & 0.0960 \\ 
		0.15 & 0.1509 & 0.1479 & 0.1517 & 0.1548 & 0.1565 & 0.1448 \\ 
		0.20 & 0.1965 & 0.1979 & 0.1967 & 0.2033 & 0.2047 & 0.1945 \\ 
		0.25 & 0.2493 & 0.2496 & 0.2441 & 0.2571 & 0.2528 & 0.2455 \\ 
		0.30 & 0.3052 & 0.2992 & 0.2949 & 0.3059 & 0.3052 & 0.2974 \\ 
		0.35 & 0.3578 & 0.3481 & 0.3496 & 0.3590 & 0.3571 & 0.3483 \\ 
		0.40 & 0.4048 & 0.3941 & 0.3958 & 0.4096 & 0.4071 & 0.4000 \\ 
		0.45 & 0.4552 & 0.4462 & 0.4435 & 0.4594 & 0.4542 & 0.4521 \\ 
		0.50 & 0.5005 & 0.4940 & 0.4954 & 0.5051 & 0.5025 & 0.5014 \\ 
		0.55 & 0.5506 & 0.5439 & 0.5458 & 0.5557 & 0.5525 & 0.5551 \\ 
		0.60 & 0.6002 & 0.5946 & 0.5977 & 0.6015 & 0.6000 & 0.6043 \\ 
		0.65 & 0.6520 & 0.6414 & 0.6460 & 0.6463 & 0.6504 & 0.6548 \\ 
		0.70 & 0.7032 & 0.6908 & 0.6988 & 0.6963 & 0.7002 & 0.7046 \\ 
		0.75 & 0.7533 & 0.7434 & 0.7494 & 0.7476 & 0.7477 & 0.7546 \\ 
		0.80 & 0.8046 & 0.7967 & 0.7986 & 0.7958 & 0.7987 & 0.8023 \\ 
		0.85 & 0.8501 & 0.8455 & 0.8526 & 0.8466 & 0.8500 & 0.8508 \\ 
		0.90 & 0.8978 & 0.8981 & 0.9005 & 0.8983 & 0.9016 & 0.8996 \\ 
		0.95 & 0.9506 & 0.9470 & 0.9507 & 0.9494 & 0.9510 & 0.9502 \\ 
			\hline
		\end{tabular}
	\end{center}
\end{table}
\begin{table}[ht]
	\caption{ Coverage Analysis of $\psun$-IS: Frequentist Coverage vs Theoretical Coverage for the Other Parameters, in the logistic regression. Different Combinations of Sample Size $n$ and Number of Parameters $p$. Number of simulations $N = 10^4$. }
	\label{IScoverage_tab_notIntercept}
	\begin{center}
		\begin{tabular}{ | c | c c | c c | c c | }
			\hline
			\multirow{2}{*}{Theoretical Coverage} & \multicolumn{2}{c|}{$n = 50$} & \multicolumn	{2}{c|}{$n = 100$} & \multicolumn{2}{c|}{$n = 200$} \\
			& $p = 500$ & $p = 1000$ & $p = 500$ & $p = 1000$ & $p = 500$ & $p = 1000$ \\ 
			\hline
			0.05 & 0.0503 & 0.0471 & 0.0476 & 0.0500 & 0.0483 & 0.0496 \\ 
			0.10 & 0.0993 & 0.0945 & 0.0942 & 0.0988 & 0.0942 & 0.0986 \\ 
			0.15 & 0.1492 & 0.1460 & 0.1420 & 0.1510 & 0.1461 & 0.1506 \\ 
			0.20 & 0.2024 & 0.1994 & 0.1916 & 0.2005 & 0.1965 & 0.1953 \\ 
			0.25 & 0.2530 & 0.2486 & 0.2431 & 0.2504 & 0.2466 & 0.2473 \\ 
			0.30 & 0.3016 & 0.2974 & 0.2951 & 0.2968 & 0.2926 & 0.3009 \\ 
			0.35 & 0.3514 & 0.3467 & 0.3446 & 0.3501 & 0.3441 & 0.3501 \\ 
			0.40 & 0.3967 & 0.3986 & 0.3943 & 0.3987 & 0.3933 & 0.3968 \\ 
			0.45 & 0.4505 & 0.4472 & 0.4388 & 0.4529 & 0.4420 & 0.4487 \\ 
			0.50 & 0.4979 & 0.4937 & 0.4906 & 0.5026 & 0.4950 & 0.4966 \\ 
			0.55 & 0.5484 & 0.5445 & 0.5395 & 0.5496 & 0.5429 & 0.5451 \\ 
			0.60 & 0.6013 & 0.5977 & 0.5914 & 0.5980 & 0.5949 & 0.5967 \\ 
			0.65 & 0.6536 & 0.6467 & 0.6417 & 0.6493 & 0.6457 & 0.6481 \\ 
			0.70 & 0.7010 & 0.6963 & 0.6945 & 0.6995 & 0.6935 & 0.6982 \\ 
			0.75 & 0.7486 & 0.7462 & 0.7447 & 0.7505 & 0.7450 & 0.7462 \\ 
			0.80 & 0.7992 & 0.7946 & 0.7974 & 0.8022 & 0.7937 & 0.7984 \\ 
			0.85 & 0.8485 & 0.8441 & 0.8445 & 0.8516 & 0.8446 & 0.8481 \\ 
			0.90 & 0.8965 & 0.8980 & 0.8972 & 0.9018 & 0.8999 & 0.8953 \\ 
			0.95 & 0.9487 & 0.9481 & 0.9489 & 0.9501 & 0.9506 & 0.9495 \\  
			\hline
		\end{tabular}
	\end{center}
\end{table}
\begin{table}[ht]
	\caption{ Coverage Analysis of $\psun$-IS: ESS in the logistic regression. Different Combinations of Sample Size $n$ and Number of Parameters $p$. Number of simulations $N = 10^4$. }
	\label{IScoverage_ESS}
	\begin{center}
		\begin{tabular}{|c c|c c c c c c|}
			\hline
			\textbf{n} & \textbf{p} & \textbf{Min} & \textbf{1st Qu.} & \textbf{Median} & \textbf{Mean} & \textbf{3rd Qu.} & \textbf{Max} \\
			\hline
			50 & 500 & 4133.61 & 4503.90 & 4554.53 & 4545.46 & 4596.64 & 4771.25 \\ 
  		50 & 1000 & 3414.52 & 3606.03 & 3640.77 & 3637.86 & 3671.86 & 3812.79 \\ 
  		100 & 500 & 3475.38 & 3859.21 & 3917.76 & 3912.14 & 3972.93 & 4201.83 \\ 
  		100 & 1000 & 3084.04 & 3272.89 & 3308.91 & 3305.91 & 3342.37 & 3487.81 \\ 
  		200 & 500 & 2487.92 & 2862.23 & 2933.14 & 2930.93 & 3004.14 & 3305.85 \\ 
  		200 & 1000 & 2529.89 & 2751.93 & 2790.23 & 2788.35 & 2827.38 & 2988.85 \\ 
			\hline
		\end{tabular}
	\end{center}
\end{table}
\begin{table}[ht]
	\caption{ Coverage Analysis of $\psun$-IS: Computational Time in seconds in the logistic regression. Different Combinations of Sample Size $n$ and Number of Parameters $p$. Number of simulations $N = 10^4$. }
	\label{IScoverage_time}
	\begin{center}
	\scalebox{0.95}{
		\begin{tabular}{|c c|c c c c c c|}
			\hline
			\textbf{n} & \textbf{p} & \textbf{Min} & \textbf{1st Qu.} & \textbf{Median} & \textbf{Mean} & \textbf{3rd Qu.} & \textbf{Max} \\
			\hline
			50 & 500 & 10.26 & 11.49 & 11.63 & 11.66 & 11.82 & 13.90 \\ 
      50 & 1000 & 14.39 & 16.10 & 16.26 & 16.29 & 16.48 & 18.51 \\ 
      100 & 500 & 21.08 & 24.27 & 24.88 & 24.97 & 25.54 & 30.59 \\ 
  		100 & 1000 & 24.52 & 27.99 & 28.48 & 28.55 & 29.04 & 32.25 \\ 
  		200 & 500 & 271.21 & 498.92 & 562.52 & 572.32 & 634.49 & 1188.48 \\ 
  		200 & 1000 & 71.37 & 101.05 & 107.51 & 108.43 & 114.77 & 171.88 \\ 
			\hline
		\end{tabular}
		}
	\end{center}
\end{table}
\subsection{Simulation Study: Polya-Gamma vs \texorpdfstring{$\boldsymbol{\psun}$}{\psun} with Small Sample Size}

Tables \ref{PGvs_pSUN_Gauss_ESSps}, \ref{PGvs_pSUN_Laplacit_ESSps}, and \ref{PGvs_pSUN_DirLapl_ESSps} report the mean of the \textit{effective sample size per second} using the Gaussian, Laplacit, and Dirichlet-Laplace prior respectively. The $\psun$-Gibbs is more efficient than competitive MCMC methods and the $\psun$-IS, when applicable, is the most efficient one and causes a dramatic improvement compared to Gibbs-based methods.

Tables \ref{PGvs_pSUN_Gauss_ESS}, \ref{PGvs_pSUN_Laplacit_ESS}, and \ref{PGvs_pSUN_DirLapl_ESS} report the mean of the \textit{effective sample size}. It is interesting to note that, using the $\psun$-Gibbs algorithm with a Gaussian prior, draws are essentially uncorrelated.

\begin{center}
	\begin{figure}
		\caption{ Polya-Gamma vs $\psun$: Average of Posterior Means for the Different Groups of $\beta$ Using the Dirichlet-Laplace Prior and Different Algorithms. Left: $\psun$-Gibbs (cross) vs UPG (bullet). Right: $\psun$-Gibbs (cross) vs PG (bullet). }
		\label{PGvspSUN_DirLaplMPM}
		\centerline{\includegraphics[scale=1]{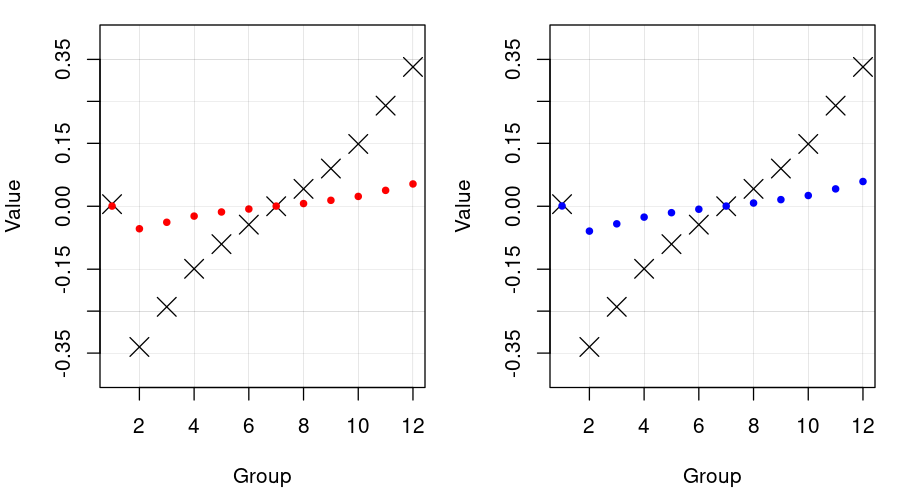}}
	\end{figure}
\end{center}

Figure \ref{PGvspSUN_DirLaplMPM} compares the average of the posterior mean under the Dirichlet-Laplace prior using the Polya-Gamma methods and the $\psun$-Gibbs.
Whereas for the other priors, differences are negligible, the Dirichlet-Laplace prior induces a clear disagreement between Polya-Gamma methods and the $\psun$-Gibbs. We believe that a convergence issue arises using the Polya-Gamma methods since all the estimates are shrunk to $0$.

\begin{table}[ht]
	\caption{ Polya-Gamma vs $\psun$: Gaussian Prior. Mean of ESS per second for the Different Sets of $\beta$ Values and Different Algorithms. Number of simulations $N = 10^4$. }
	\label{PGvs_pSUN_Gauss_ESSps}
	\begin{center}
		\begin{tabular}{| c | c c c c |}
			\hline
			 & \textbf{UPG} & \textbf{PG} & \textbf{pSUN-Gibbs} & \textbf{pSUN-IS} \\ 
			 \hline
			 Intercept & 0.42 & 1.51 & 44.34 & 374.41 \\ 
			 $\beta = \Lambda^{-1}(0.05)$ & 10.57 & 19.29 & 44.32 & 374.41 \\ 
			 $\beta = \Lambda^{-1}(0.10)$ & 10.55 & 19.26 & 44.32 & 374.41 \\ 
			 $\beta = \Lambda^{-1}(0.20)$ & 10.53 & 19.28 & 44.31 & 374.41 \\ 
			 $\beta = \Lambda^{-1}(0.30)$ & 10.52 & 19.25 & 44.31 & 374.41 \\ 
			 $\beta = \Lambda^{-1}(0.40)$ & 10.56 & 19.27 & 44.32 & 374.41 \\ 
			 $\beta = \Lambda^{-1}(0.50)$ & 10.55 & 19.27 & 44.32 & 374.41 \\ 
			 $\beta = \Lambda^{-1}(0.60)$ & 10.53 & 19.26 & 44.32 & 374.41 \\ 
			 $\beta = \Lambda^{-1}(0.70)$ & 10.56 & 19.28 & 44.32 & 374.41 \\ 
			 $\beta = \Lambda^{-1}(0.80)$ & 10.55 & 19.25 & 44.32 & 374.41 \\ 
			 $\beta = \Lambda^{-1}(0.90)$ & 10.54 & 19.27 & 44.31 & 374.41 \\ 
			 $\beta = \Lambda^{-1}(0.95)$ & 10.59 & 19.25 & 44.33 & 374.41 \\ 
			\hline
		\end{tabular}
	\end{center}
\end{table}
\begin{table}[ht]
	\caption{ Polya-Gamma vs $\psun$: Laplacit Prior. Mean of ESS per second for the Different Groups of $\beta$ and Algorithms. Number of simulations $N = 10^4$. }
	\label{PGvs_pSUN_Laplacit_ESSps}
	\begin{center}
		\begin{tabular}{| c | c c c |}
			\hline
			 & \textbf{UPG} & \textbf{PG} & \textbf{$\bm \psun$-Gibbs} \\ 
			 \hline
  		 Intercept & 0.36 & 1.30 & 24.80 \\ 
  		 $\beta = \Lambda^{-1}(0.05)$ & 7.30 & 13.89 & 26.15 \\ 
  		 $\beta = \Lambda^{-1}(0.10)$ & 7.34 & 13.97 & 26.35 \\ 
  		 $\beta = \Lambda^{-1}(0.20)$ & 7.34 & 14.06 & 26.52 \\ 
  	   $\beta = \Lambda^{-1}(0.30)$ & 7.38 & 14.09 & 26.58 \\ 
  		 $\beta = \Lambda^{-1}(0.40)$ & 7.35 & 14.08 & 26.60 \\ 
  		 $\beta = \Lambda^{-1}(0.50)$ & 7.39 & 14.10 & 26.64 \\ 
  		 $\beta = \Lambda^{-1}(0.60)$ & 7.38 & 14.10 & 26.61 \\ 
  		 $\beta = \Lambda^{-1}(0.70)$ & 7.39 & 14.09 & 26.58 \\ 
  		 $\beta = \Lambda^{-1}(0.80)$ & 7.36 & 14.04 & 26.50 \\ 
  		 $\beta = \Lambda^{-1}(0.90)$ & 7.34 & 13.97 & 26.36 \\ 
  		 $\beta = \Lambda^{-1}(0.95)$ & 7.32 & 13.86 & 26.12 \\ 
			\hline
		\end{tabular}
	\end{center}
\end{table}
\begin{table}[ht]
	\caption{ Polya-Gamma vs $\psun$: Dirichlet-Laplace Prior. Mean of ESS per second for the Different Groups of $\beta$ and Algorithms. Number of simulations $N = 10^5$. }
	\label{PGvs_pSUN_DirLapl_ESSps}
	\begin{center}
		\begin{tabular}{| c | c c c |}
			\hline
			 & \textbf{UPG} & \textbf{PG} & \textbf{$\bm \psun$-Gibbs} \\ 
			 \hline
			 Intercept & 0.78 & 1.95 & 11.12 \\ 
			 $\beta = \Lambda^{-1}(0.05)$ & 0.46 & 1.23 & 7.83 \\ 
			 $\beta = \Lambda^{-1}(0.10)$ & 0.48 & 1.27 & 8.06 \\ 
			 $\beta = \Lambda^{-1}(0.20)$ & 0.48 & 1.29 & 8.21 \\ 
			 $\beta = \Lambda^{-1}(0.30)$ & 0.49 & 1.31 & 8.28 \\ 
			 $\beta = \Lambda^{-1}(0.40)$ & 0.49 & 1.31 & 8.32 \\ 
			 $\beta = \Lambda^{-1}(0.50)$ & 0.49 & 1.32 & 8.33 \\ 
			 $\beta = \Lambda^{-1}(0.60)$ & 0.49 & 1.32 & 8.34 \\ 
			 $\beta = \Lambda^{-1}(0.70)$ & 0.49 & 1.31 & 8.31 \\ 
			 $\beta = \Lambda^{-1}(0.80)$ & 0.49 & 1.30 & 8.24 \\ 
			 $\beta = \Lambda^{-1}(0.90)$ & 0.47 & 1.27 & 8.05 \\ 
			 $\beta = \Lambda^{-1}(0.95)$ & 0.46 & 1.24 & 7.87 \\
			\hline
		\end{tabular}
	\end{center}
\end{table}
\begin{table}[ht]
	\caption{ Polya-Gamma vs $\psun$: Gaussian Prior. Mean of ESS for the Different Groups of $\beta$ and Algorithms. Number of simulations $N = 10^4$. }
	\label{PGvs_pSUN_Gauss_ESS}
	\begin{center}
		\begin{tabular}{| c | c c c c |}
			\hline
			 & \textbf{UPG} & \textbf{PG} & \textbf{$\bm \psun$-Gibbs} & \textbf{$\bm \psun$-IS} \\ 
			 \hline
			 Intercept & 103.39 & 420.35 & $>$9999.99 & 4579.48 \\ 
			 $\beta = \Lambda^{-1}(0.05)$ & 2592.88 & 5381.69 & $>$9999.99 & 4579.48 \\ 
			 $\beta = \Lambda^{-1}(0.10)$ & 2588.63 & 5373.54 & $>$9999.99 & 4579.48 \\ 
			 $\beta = \Lambda^{-1}(0.20)$ & 2583.95 & 5380.46 & $>$9999.99 & 4579.48 \\ 
			 $\beta = \Lambda^{-1}(0.30)$ & 2580.51 & 5371.88 & $>$9999.99 & 4579.48 \\ 
			 $\beta = \Lambda^{-1}(0.40)$ & 2590.55 & 5374.92 & $>$9999.99 & 4579.48 \\ 
			 $\beta = \Lambda^{-1}(0.50)$ & 2589.06 & 5376.31 & $>$9999.99 & 4579.48 \\ 
			 $\beta = \Lambda^{-1}(0.60)$ & 2583.75 & 5373.65 & $>$9999.99 & 4579.48 \\ 
			 $\beta = \Lambda^{-1}(0.70)$ & 2590.87 & 5378.66 & $>$9999.99 & 4579.48 \\ 
			 $\beta = \Lambda^{-1}(0.80)$ & 2588.97 & 5371.48 & $>$9999.99 & 4579.48 \\ 
			 $\beta = \Lambda^{-1}(0.90)$ & 2585.55 & 5376.66 & $>$9999.99 & 4579.48 \\ 
			 $\beta = \Lambda^{-1}(0.95)$ & 2599.04 & 5370.28 & $>$9999.99 & 4579.48 \\ 
			\hline
		\end{tabular}
	\end{center}
\end{table}
\begin{table}[ht]
	\caption{ Polya-Gamma vs $\psun$: Laplacit Prior. Mean of ESS for the Different Groups of $\beta$ and Algorithms. Number of simulations $N = 10^4$. }
	\label{PGvs_pSUN_Laplacit_ESS}
	\begin{center}
		\begin{tabular}{| c | c c c |}
			\hline
			 & \textbf{UPG} & \textbf{PG} & \textbf{$\bm \psun$-Gibbs} \\ 
			 \hline
			 Intercept & 129.75 & 475.24 & 8430.86 \\ 
			 $\beta = \Lambda^{-1}(0.05)$ & 2597.43 & 5088.88 & 8889.60 \\ 
			 $\beta = \Lambda^{-1}(0.10)$ & 2613.34 & 5119.66 & 8959.38 \\ 
			 $\beta = \Lambda^{-1}(0.20)$ & 2613.97 & 5152.85 & 9015.70 \\ 
			 $\beta = \Lambda^{-1}(0.30)$ & 2625.40 & 5161.63 & 9038.21 \\ 
			 $\beta = \Lambda^{-1}(0.40)$ & 2617.74 & 5160.85 & 9043.51 \\ 
			 $\beta = \Lambda^{-1}(0.50)$ & 2630.42 & 5167.49 & 9055.93 \\ 
			 $\beta = \Lambda^{-1}(0.60)$ & 2628.62 & 5165.18 & 9047.44 \\ 
			 $\beta = \Lambda^{-1}(0.70)$ & 2630.32 & 5164.77 & 9036.62 \\ 
			 $\beta = \Lambda^{-1}(0.80)$ & 2619.68 & 5145.53 & 9009.29 \\ 
			 $\beta = \Lambda^{-1}(0.90)$ & 2611.95 & 5118.46 & 8962.14 \\ 
			 $\beta = \Lambda^{-1}(0.95)$ & 2604.17 & 5079.74 & 8880.55 \\ 
			\hline
		\end{tabular}
	\end{center}
\end{table}
\begin{table}[ht]
	\caption{ Polya-Gamma vs $\psun$: Dirichlet-Laplace Prior. Mean of ESS for the Different Groups of $\beta$ and Algorithms. Number of simulations $N = 10^5$. }
	\label{PGvs_pSUN_DirLapl_ESS}
	\begin{center}
		\begin{tabular}{| c | c c c |}
			\hline
			 & \textbf{UPG} & \textbf{PG} & \textbf{$\bm \psun$-Gibbs} \\ 
			 \hline
			 Intercept & 6616.88 & 10516.49 & 43126.76 \\ 
			 $\beta = \Lambda^{-1}(0.05)$ & 3926.72 & 6635.02 & 30247.73 \\ 
  		 $\beta = \Lambda^{-1}(0.10)$ & 4029.54 & 6847.65 & 31106.55 \\ 
  		 $\beta = \Lambda^{-1}(0.20)$ & 4093.37 & 6990.94 & 31687.74 \\ 
  		 $\beta = \Lambda^{-1}(0.30)$ & 4131.65 & 7054.91 & 31955.14 \\ 
  		 $\beta = \Lambda^{-1}(0.40)$ & 4149.70 & 7098.72 & 32110.49 \\ 
  		 $\beta = \Lambda^{-1}(0.50)$ & 4166.75 & 7109.17 & 32167.66 \\ 
  		 $\beta = \Lambda^{-1}(0.60)$ & 4167.96 & 7112.73 & 32172.67 \\ 
  		 $\beta = \Lambda^{-1}(0.70)$ & 4137.10 & 7081.33 & 32058.97 \\ 
  		 $\beta = \Lambda^{-1}(0.80)$ & 4113.74 & 7003.26 & 31798.48 \\ 
  		 $\beta = \Lambda^{-1}(0.90)$ & 4016.79 & 6840.76 & 31072.15 \\ 
  		 $\beta = \Lambda^{-1}(0.95)$ & 3928.75 & 6695.49 & 30384.44 \\ 
			\hline
		\end{tabular}
	\end{center}
\end{table}
\subsection{Simulation Study: a Probit Model with Sparse Parameters}

Table\ref{probitSparse_ESS} reports the \textit{effective sample size}. Table \ref{probitSparse_times} reports summaries of computational time in seconds for the different priors.

\begin{table}[ht]
	\caption{ Probit Model with Sparse Parameters: Mean of ESS for the Different Groups of $\beta$ and Priors. Number of simulations $N = 10^4$ for Gaussian and Laplacit and $N = 10^5$ for Dirichlet-Laplace. }
	\label{probitSparse_ESS}
	\begin{center}
		\begin{tabular}{| c | c c c |}
			\hline
			& \textbf{Gaussian} & \textbf{Laplacit} & \textbf{Dirichlet-Laplace} \\ 
			\hline 
  		Intercept & 10000.00 & 8222.25 & 40723.28 \\ 
  		$\Phi^{-1}(0.05)$ & 10000.00 & 8560.60 & 26573.48 \\ 
  		$\Phi^{-1}(0.10)$ & 10000.00 & 8746.52 & 28633.32 \\ 
  		$\Phi^{-1}(0.20)$ & 10000.00 & 8920.11 & 30718.95 \\ 
  		$\Phi^{-1}(0.30)$ & 10000.00 & 8996.87 & 31574.69 \\ 
  		$\Phi^{-1}(0.40)$ & 10000.00 & 9043.80 & 31929.77 \\ 
  		$\Phi^{-1}(0.50)$ & 10000.00 & 9046.28 & 32108.00 \\ 
  		$\Phi^{-1}(0.60)$ & 10000.00 & 9035.04 & 31817.41 \\ 
  		$\Phi^{-1}(0.70)$ & 10000.00 & 8997.00 & 31557.77 \\ 
  		$\Phi^{-1}(0.80)$ & 10000.00 & 8898.38 & 30569.49 \\ 
  		$\Phi^{-1}(0.90)$ & 10000.00 & 8746.65 & 28538.98 \\ 
  		$\Phi^{-1}(0.95)$ & 10000.00 & 8546.51 & 26735.17 \\  
			\hline
		\end{tabular}
	\end{center}
\end{table}
\begin{table}[ht]
	\caption{ Probit Model with Sparse Parameters: Summaries of Computational Time in seconds for the Different Priors. Number of simulations $N = 10^4$ for Gaussian and Laplacit and $N = 10^5$ for Dirichlet-Laplace. }
	\label{probitSparse_times}
	\begin{center}
	\scalebox{0.90}{
			\begin{tabular}{| c | c c c c c c |}
				\hline
				\textbf{Prior} & \textbf{Min} & \textbf{1st Qu.} & \textbf{Median} & \textbf{Mean} & \textbf{3rd Qu.} & \textbf{Max} \\
				\hline
				Gaussian & 1.55 & 1.74 & 1.81 & 1.85 & 1.89 & 3.63 \\ 
  			    Laplacit & 60.91 & 63.57 & 64.09 & 64.54 & 64.96 & 74.76 \\ 
  			    Dirichlet-Laplace & 3258.98 & 3626.25 & 3741.17 & 3774.31 & 3923.47 & 4426.62 \\ 
				\hline
			\end{tabular}
			}
		\end{center}
\end{table}
\subsection{Real Data Analysis: Parameters Estimation for the \textit{Cancer SAGE Dataset}}

\begin{center}
  \begin{figure}
		\caption{ Cancer SAGE Example, Logit Model: Posterior Means of the 516 $\beta$ Coefficients Plus The Intercept $\beta_1$. Dirichlet-Laplace Prior with Different Algorithms. Left: UPG. Center: PG. Right: $\psun$-Gibbs. }
 		\label{CancerSAGE_logitEst_DirLapl}
 		\centerline{\includegraphics[scale=1]{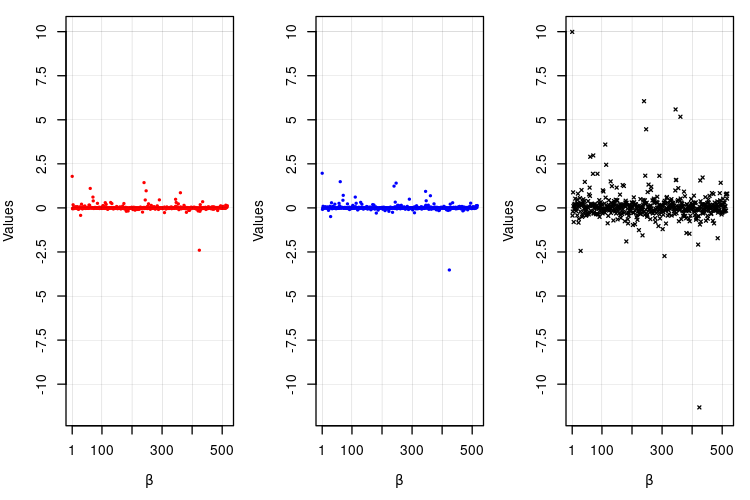}}
 	\end{figure}
\end{center}

\begin{center}
  \begin{figure}
		\caption{ Cancer SAGE Example, Logit Model: Posterior Medians of the 516 $\beta$ Coefficients Plus The Intercept $\beta_1$. Cauchy Prior with Different Algorithms. Left: UPG. Center: PG. Right: $\psun$-Gibbs. }
 		\label{CancerSAGE_medLogitCauchy}
 		\centerline{\includegraphics[scale=1]{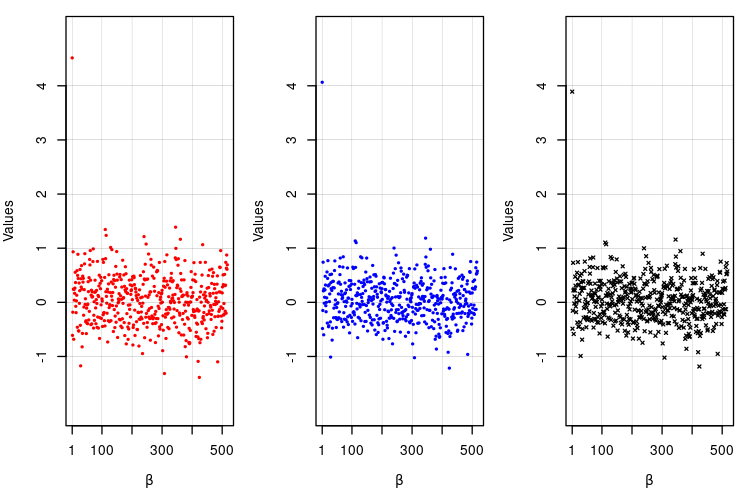}}
 	\end{figure}
\end{center}

Figure \ref{CancerSAGE_logitEst_DirLapl} reports the posterior means for the logit case with the Dirichlet-Laplace prior under the use of different algorithms. Figure \ref{CancerSAGE_medLogitCauchy} reports the posterior medians for the logit model combined with the Cauchy prior for the different algorithms.
Table \ref{CancerSAGE_logitESS} reports summaries of ESS for the logit case with different priors and different algorithms. Table \ref{CancerSAGE_probitESS} reports summaries of ESS for the probit case with different priors. Table \ref{CancerSAGE_times} reports the computational times in seconds for all combinations of links, priors, and algorithms.


\begin{table}[ht]
	\caption{ Cancer SAGE Example, Logit Model: Summaries of ESS using different algorithms and priors. Number of simulations $N = 10^5$. } 
	\label{CancerSAGE_logitESS}
	\begin{center}
	\scalebox{0.95}{
		\begin{tabular}{| c c | c c c c c c |}
			\hline
			\textbf{Algorithm} & \textbf{Prior} & \textbf{Min} & \textbf{1st Qu.} & \textbf{Median} &  \textbf{Mean} & \textbf{3rd Qu.} & \textbf{Max} \\
			\hline
			UPG & Gaussian & 589.79 & 4894.55 & 7146.74 & 8759.66 & 10939.96 & 46370.16 \\ 
			PG & Gaussian & 3166.46 & 35089.79 & 40551.23 & 42043.24 & 47375.16 & 84783.86 \\ 
			$\psun$-Gibbs & Gaussian & 95987.47 & $>$99999.99 & $>$99999.99 & 99790.01 & $>$99999.99 & $>$99999.99 \\ 
			$\psun$-IS & Gaussian & 39836.20 & 39836.20 & 39836.20 & 39836.20 & 39836.20 & 39836.20 \\ 
			\hdashline
			UPG & Laplacit & 321.51 & 5254.60 & 7299.22 & 8819.87 & 10730.65 & 38771.20 \\ 
			PG & Laplacit & 1646.02 & 31745.38 & 37179.79 & 38335.75 & 44144.48 & 78874.34 \\ 
			$\psun$-Gibbs & Laplacit & 41327.58 & 83157.00 & 91992.76 & 88520.78 & 97323.73 & $>$99999.99 \\ 
			\hdashline
			UPG & Dirichlet-Laplace & 219.50 & 1587.87 & 3226.43 & 3962.58 & 5791.68 & 17891.26 \\ 
			PG & Dirichlet-Laplace & 326.41 & 2422.83 & 5201.40 & 6505.77 & 9703.00 & 29107.47 \\ 
			$\psun$-Gibbs & Dirichlet-Laplace & 1925.68 & 15990.32 & 26856.47 & 28803.47 & 40560.04 & 64053.01 \\
			\hline
		\end{tabular}
	}
	\end{center}
\end{table}
\begin{table}[ht]
	\caption{ Cancer SAGE Example, Probit Model: Summaries of ESS using different priors. Number of simulations $N = 10^5$. }
	\label{CancerSAGE_probitESS}
	\begin{center}
		\begin{tabular}{| c | c c c c c c |}
			\hline
			\textbf{Prior} & \textbf{Min} & \textbf{1st Qu.} & \textbf{Median} & \textbf{Mean} & \textbf{3rd Qu.} & \textbf{Max} \\
			\hline
			Gaussian & 100000.00 & 100000.00 & 100000.00 & 100000.00 & 100000.00 & 100000.00 \\ 
  		Laplacit & 42047.62 & 83148.71 & 92058.81 & 88477.74 & 97128.17 & 102140.93 \\ 
  		Dirichlet-Laplace & 2012.51 & 16275.66 & 26963.41 & 28614.40 & 40521.48 & 65296.70 \\ 
			\hline
		\end{tabular}
	\end{center}
\end{table}
\begin{table}[ht]
	\caption{ Cancer SAGE Example: Computational Time (Seconds) for Different Links, Algorithms, and Priors ($\sun$-rng is the sampler for the $\sun$ distribution). Number of simulations $N = 10^5$. }
	\label{CancerSAGE_times}
	\begin{center}
	\scalebox{0.90}{
			\begin{tabular}{|c c c | c |}
				\hline
				\textbf{Link} & \textbf{Algorithm} & \textbf{Prior} & \textbf{Time} \\
				\hline
				\multirow{10}{*}{Logit} & UPG & Gaussian & 2802.10 \\ 
				 & PG & Gaussian & 2815.15 \\ 
				 & $\psun$-Gibbs & Gaussian & 3235.48 \\ 
				 & $\psun$-IS & Gaussian & 189.87 \\ 
				 & UPG & Laplacit & 3894.09 \\ 
				 & PG & Laplacit & 3865.85 \\ 
				 & $\psun$-Gibbs & Laplacit & 4415.48 \\ 
				 & UPG & Dirichlet-Laplace & 8409.55 \\ 
				 & PG & Dirichlet-Laplace & 5228.17 \\ 
				 & $\psun$-Gibbs & Dirichlet-Laplace & 23720.55 \\
				 \hdashline 
				 \multirow{3}{*}{Probit} & $\sun$-rng & Gaussian & 54.21 \\ 
				 & $\psun$-Gibbs & Laplacit & 4415.30 \\ 
				 & $\psun$-Gibbs & Dirichlet-Laplace & 23742.32 \\ 
				\hline
			\end{tabular}
			}
		\end{center}
\end{table}